\documentclass[aps,pre,superscriptaddress,floatfix]{revtex4}
\usepackage[dvips]{graphics}
\usepackage{graphicx}
\usepackage{amsfonts}
\usepackage{amssymb}
\usepackage{amsmath}
\usepackage{subfigure}

\begin{document}

\title{Phase-ordering kinetics on graphs}

\author{R. Burioni}
\affiliation{Dipartimento di Fisica and INFN, Universit\`a di Parma, 
Parco Area delle Scienze 7/A, I-423100 Parma, Italy}
\author{D. Cassi}
\affiliation{Dipartimento di Fisica and INFN, Universit\`a di Parma, 
Parco Area delle Scienze 7/A, I-423100 Parma, Italy}
\author{F.Corberi}
\affiliation{Dipartimento di Fisica ``E.R.Caianiello'' and CNR-INFM, Universit\`a 
di Salerno, 84081 Baronissi (Salerno), Italy}
\author{A. Vezzani}
\affiliation{Dipartimento di Fisica and CNR-INFM, Universit\`a di Parma, 
Parco Area delle Scienze 7/A, I-423100 Parma, Italy}

 \date{\today}

\begin{abstract}

We study numerically the phase-ordering kinetics following a temperature quench 
of the Ising model with single spin flip dynamics on a class of graphs, 
including geometrical fractals and random fractals, such as the percolation cluster. 
For each structure we discuss the scaling properties and compute the dynamical exponents.
We show that the exponent $a_\chi$ for the integrated response function, at variance with
all the other exponents, is independent on temperature and on the presence of pinning.
This universal charachter suggests a strict
relation between $a_\chi$ and the topological properties of the networks, 
in analogy to what observed on regular lattices. 

\end{abstract}


\maketitle

\draft
\def\be{\begin{equation}}
\def\ee{\end{equation}}
\def\bfi{\begin{figure}}
\def\efi{\end{figure}}
\def\bea{\begin{eqnarray}}
\def\eea{\end{eqnarray}}

\section{Introduction}

After quenching a system from an high temperature disordered state to
an ordered phase with broken ergodicity, a phase-ordering kinetics
is observed, characterized by coarsening of quasi-equilibrated domains
with a typical size $L(t)$. 
Although a first-principle theory of phase-ordering is presently
lacking, for systems defined on homogeneous lattices
a substantial comprehension of the dynamics has been achieved by
means of exact solutions of soluble cases, approximate theories 
and numerical simulations~\cite{Bray94}. 
While the inner of domains is basically in equilibrium at the
quench temperature, the off-equilibrium behavior is provided
by the slow evolution of their boundaries. As a consequence,
at each time $s$, for space separations $r \ll L(s)$ or for
time separations $t-s\ll s$ intra-domains quasi-equilibrium 
properties are probed, while for $r \gg L(s)$ or $t-s\gg s$
one explores inter-domains properties, where the aging behavior
is observed. Accordingly, the
correlation of the order parameter $\langle \sigma _i(t)\sigma _j(s)\rangle $
between sites $i,j$ at times $s,t$ can be expressed as the
sum of two terms
\be
{\cal C}_{ij}(t,s)={\cal C}^{st}_{ij}(t-s)+{\cal C}^{ag}_{ij}(t,s).
\label{1}
\ee 
The first term describes the equilibrium 
contribution provided by the interior of 
domains while the second contains the non-equilibrium information.
Analogously, also the integrated response function, or zero field 
cooled magnetization,  measured on site $i$ 
at time $t$
after a perturbation has switched on in $j$ from time $s$ onwards, 
takes an analogous addictive form
\be
\chi_{ij}(t,s)=\chi^{st}_{ij}(t-s)+\chi^{ag}_{ij}(t,s).
\label{1a}
\ee 
On regular lattices, due to space homogeneity and isotropy, 
correlation and response function depend only
on the distance $r$ between $i$ and $j$. One has, therefore,
${\cal C}_{ij}(t,s)={\cal C}(r,t,s)$, and similarly for $\chi_{ij}(t,s)$.

At the heart of the non-equilibrium behavior is the 
dynamical scaling symmetry, a self-similarity where
time acts solely as a length rescaling. When scaling
holds, the states sequentially entered by the system are statistically  
equivalent provided lengths are measured in units of
the characteristic size $L(t)$ of ordered domains.
All the time dependence must enter through $L(t)$,
and the aging parts in Eqs.~(\ref{1},\ref{1a}) take a scaling form
in terms of rescaled variables~\cite{Furukawa89}
$x=r/L(s)$ and $y=L(t)/L(s)$  
\be
{\cal C}^{ag}(r,t,s)=\tilde {\cal C}(x,y),
\label{2}
\ee
\be
\chi ^{ag}(r,t,s)=s^{-a_\chi}\tilde \chi (x,y).
\label{2a}
\ee
The characteristic length grows according to a power law $L(t)\sim t^{1/z}$.
Non-equilibrium exponents such as $a_\chi,z$, are expected to be universal. Namely,
they depend only on a restrict set of parameters, such as space dimensionality
and number of components ${\cal N}$ of the order parameter, on the conservation
laws of the dynamics and are independent on temperature. 

In defiance of this basic comprehension in homogeneous systems, 
our understanding of phase-ordering on inhomogeneous structures is particularly
poor, although examples, ranging 
from disordered materials, to percolation clusters, glasses, polymers, 
and biomolecules, may be abundantly found in physics, economics, chemistry and biology~\cite{interd}. 

\begin{figure}
    \centering
   \rotatebox{0}{\resizebox{.7\textwidth}{!}{\includegraphics{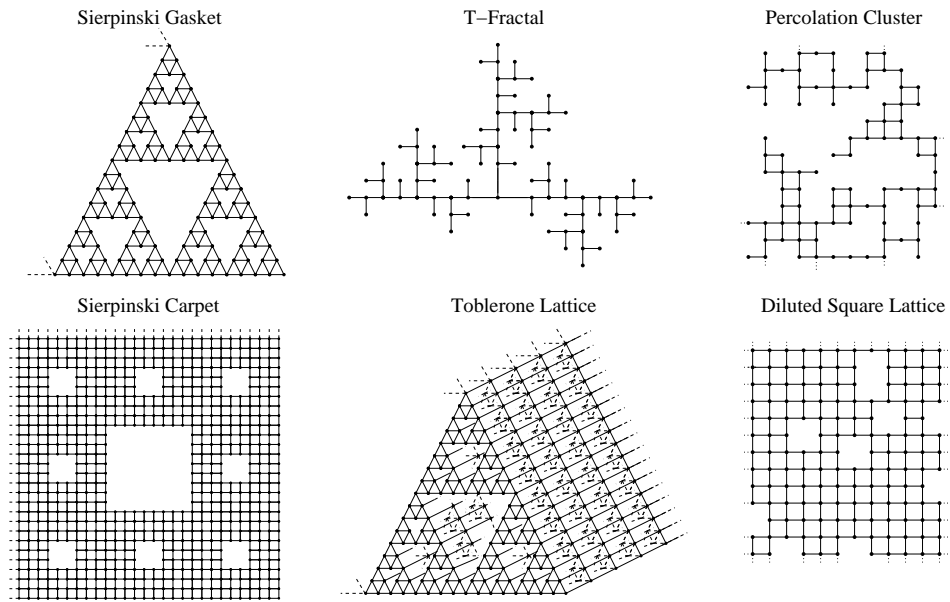}}}
    \caption{The structures considered in the paper.}
\label{strutture}
\end{figure}

In this paper we study the phase-ordering kinetics 
of the Ising model with non-conserved order parameter
on some {\it physical graphs}~\cite{rassegna}, 
namely networks with the appropriate topological features to
represent real physical structures. 
These structures are constrained to be embeddable in a finite dimensional space
and to have bounded coordination number.
Among these, we consider random fractals, i.e. percolation clusters,
and geometrical fractals, such as the Sierpinski gasket and carpet and others
(see all the cases considered in Fig.~\ref{strutture}).
We discuss the results of numerical simulations and compare them  with the predictions provided
by a large-${\cal N}$ model, where exact calculations can
be carried out~\cite{Bettolo97,prlnostro}. In the large-${\cal N}$ model the general framework of 
scaling behavior is maintained on generic graphs, and the exponents
depend on the topology of the network only through the fractal dimension $d_f$ and
the spectral dimension $d_s$, a large scale parameter encoding the relevant topology.
In particular, one finds $z=2d_f/d_s$ and $a_\chi =(d_s-2)/2$. 
This general framework provided by the soluble model seems to be quite
unrepresentative of the real situation in scalar systems, as observed in~\cite{Bettolo98}
and in our simulations.
A basic feature missed is the fundamental role played by activated processes 
on inhomogeneous structures.
Actually, on homogeneous structures the temperature is an irrelevant parameter,
in the sense of the renormalization group~\cite{Bray94}. Long-time,
large-scale properties do not depend on the quench temperature in the whole
low temperature phase. Exponents related to the aging
behavior are then independent of the thermal noise. 
On the other hand, structure inhomogeneities exert a pinning
force on interfaces or other topological defects, whose further evolution
can only be mediated by activated processes. Activated processes may
be present in non disordered models on regular lattices as well, where they sometimes play a fundamental
role, but without changing large-scale, long-time properties. 
On fractal sets, on the other hand, pinning forces are exerted on
all lengthscales and contribute to the aging behavior. 
These forces causes a stop-and-go behavior hindering the power laws and
preventing a straightforward definition of the exponents. 
At relatively high temperatures, where pinning barriers are more easily surmounted,
slip-stick effects are less severe. In these cases the general scaling
scheme~(\ref{2},\ref{2a}) can be investigated and the non equilibrium exponents 
can be defined. It turns out, however, that pinning forces are still subtly at work, 
making exponents temperature dependent.

A notable exception in this vague scenario is represented by the response function
exponent $a_\chi$. Interestingly, at variance with all the other exponents, 
its  value only depends on the structure considered, 
is independent on temperature and on the presence of stick-slip intermittency.
This universal character
calls for a strict and direct relation between $a_\chi$ and precise topological properties
of the network, bypassing any microscopic dynamical mechanism. 
We develop an argument showing the relation between this exponent
and fundamental equilibrium properties. In particular, $a_\chi$
must take a positive value whenever the statistical model on the network considered has a 
phase transition at a finite critical temperature $T_c$,
whereas $a_\chi=0$ on structures with $T_c=0$. 
The same picture is provided
by the solution of the large-${\cal N}$ model on graphs. 
This is what we find, with good accuracy, in the simulation of 
discrete symmetry models on all the inhomogeneous structures considered.
We measure $a_\chi=0$ or $a_\chi>0$ whenever structures with $T_c=0$ or
$T_c>0$ are considered.
Given the fundamental 
role of topological and connectivity
properties in determining equilibrium and critical behavior~\cite{aharony}
our results provide an evidence for some relationship 
between non-equilibrium 
kinetics and large scale topology on general networks
and suggest that the same topological 
features of graphs
determine critical behavior and non-equilibrium exponent $a_\chi$ 
during phase ordering \cite{prlnostro}. 

This paper is organized as follows: In Sec.~\ref{model} we introduce the 
Ising model that will be considered in the simulations. We also introduce
the basic observables, and discuss the numerical techniques.
In Sec.\ref{structures} we discuss the outcome of the numerical simulations
on different structures. 
In particular, subsection~\ref{phasetr}
is devoted to structures with a finite $T_c$, namely 
the percolation cluster above $p_c$, the Sierpinski carpet and
toblerone lattices, while subsection~\ref{phasetr}
deals with structures with $T_c=0$: The percolation cluster above $p_c$,
the Sierpinski gasket and the T-fractal. 
In Sec.~\ref{response} we discuss the value of the exponent $a_\chi$
in all the cases considered, and the form of the fluctuation-dissipation plots.
Sec.~\ref{concl} contains a final discussion and the conclusions. 

\section{Model and observables} \label{model}

The Ising model is defined by the Hamiltonian
\be
H[\sigma ]=-J\sum _{<ij>}\sigma _i \sigma _j,
\label{hamiltonian}
\ee
where $\sigma _i =\pm 1$ is a unitary spin and $<ij>$ are
nearest neighbours on a graph.

The dynamics is introduced by randomly choosing a single spin 
and updating it with Metropolis transition rate
\be
w([\sigma]\to [\sigma'])=\frac{1}{2}\min \left [1, \exp (-\Delta E/T)\right ].
\label{metropolis} 
\ee
Here $[\sigma]$ and $[\sigma']$ are the spin configurations before and
after the move, and
\be
\Delta E=H[\sigma']-H[\sigma].
\label{deltaE}
\ee

We consider a system of $N$ spins initially prepared in an
high temperature uncorrelated state and then quenched, at time $t=0$,
to a low final temperature $T_f$.

As already discussed, the dynamics of the spins in the bulk of domains
provide the equilibrium contribution while what is left over
accounts for the aging behavior. 
Since equilibrium dynamics is well understood, the stationary parts
in Eqs.~(\ref{1},\ref{1a}) are generally well known. 
In particular, at $T_f=0$
equilibrium dynamics is frozen and there are no stationary contributions.  
On the other hand, much interest is focused on the aging terms.
These can be isolated by subtracting the stationary parts computed
in equilibrium, from the whole quantity measured in the
phase-ordering stage. 
However it is computationally much more efficient
to resort to a different method. This amounts to study a modified system where 
$T_f$ in the transition rate~(\ref{metropolis})
is set equal to zero if the spin to be updated
belongs to the bulk, namely if it is aligned with all its neighbors. 
Since the bulk degrees of freedom, which alone contribute to the stationary parts,
feel $T_f=0$, and equilibrium dynamics is frozen at $T_f=0$,
by computing observables with this modified dynamics one isolates
the aging term leaving other properties of the dynamics 
unchanged~\cite{Corberiresp2d}. This modified dynamics
with no bulk flips (NBF) will be used in the following. 
The NBF has also the useful property 
of shifting $T_c$ towards higher values~\cite{olivera}.
As we will discuss below, this fact represents a formidable advantage
on inhomogeneous graphs.

Let us discuss the case of graphs where a 
phase transition occurs at a finite critical temperature $T_c$.
We denote these as systems of class I. 
For these structures we can consider quenches to finite
temperatures $T_f<T_c$.
The characteristic size $L(t)$ then grows until it becomes
comparable with the system size and the new equilibrium state
at $T_f$ is globally attained. We always consider
sufficiently large systems to prevent equilibration
on the simulated timescales. The dynamics in this regime is
equivalent to that of an infinite system, for which the
final equilibrium state is never reached and $L(t)$ keeps growing
indefinitely. By analogy with regular lattices, in the late stage
one might expect the power law 
\be
L(t)\sim t^{\frac{1}{z}},
\label{ldit}
\ee
although some caveats will be discussed in the following.  
On regular lattices the value $z=2$ is found quite
generally. In the solution of the large-${\cal N}$ model on graphs~\cite{Bettolo97}
or in the framework of approximate theories~\cite{Bettolo98}
one finds $z=2d_f/d_s$,
but such a relation is not expected for scalar systems.

Now we turn to systems of class II, namely those systems for
which, with standard dynamics, $T_c=0$.
For every finite
$T_f$ the system eventually equilibrate to a state with a finite coherence length
$\xi (T_f)$, which diverges in the $T_f\to 0$ limit. If the temperature is sufficiently
low, an interrupted phase-ordering is observed until $L(t)$ becomes comparable
with $\xi (T_f)$ and equilibration occurs. The phase-ordering phenomenon
can be widened at will by decreasing $T_f$, and the scaling behavior with
Eq.~(\ref{ldit}) can be studied.
Quite generally, however, on inhomogeneous structures one cannot set $T_f=0$ 
directly, because this would freeze the
dynamics due to pinning effects.
For the same reason, also very low temperatures are not numerically accessible, since
it takes an exponentially long time to surmount pinning barriers. 
From the numerical point of view, then, one has to find a reasonable
compromise between two contrasting issues. Namely, $T_f$ must be sufficiently low
in order to have a wide scaling regime preceding equilibration, but also
sufficiently large to allow pinning barriers being overtoped. 
This may be in some cases impossible. 
Fortunately, the use of the NBF dynamics simplifies considerably the problem.
As we have already pointed out, with NBF the critical temperature is raised.
In the case of system where $T_c=0$ with standard dynamics, 
the effect of NBF is to move $T_c$ from zero to some finite value. Then
one has a whole low-temperature region where phase-ordering occurs
asymptotically without being interrupted by equilibration, much 
alike in systems of class I. This allows to study the scaling behavior also
in this case. For all the reasons discussed above, in this paper
we will always present results obtained with NBF dynamics.

Generally speaking, when scaling holds
the characteristic length $L(t)$ can be estimated from the
knowledge of the two-points equal time correlation function,
obtained letting $t=s$ in Eq.~(\ref{1})
\be
G_{i,j}^{ag}(t)=\langle \vec \sigma_i(t) \cdot \vec \sigma_j(t) \rangle
\label{gdir}
\ee
where $\langle \dots \rangle$
means an ensemble average, namely taken over different initial conditions
and thermal histories.
In homogeneous systems according to Eq.~(\ref{2})
one has
\be
G_{i,j}^{ag}(t)=G^{ag}(r,t)=g(x),
\label{gdirscal}
\ee
where $x=r/L(t)$. 
For scalar order parameter with sharp interfaces, a short distance behavior
($x\ll 1$) of the type $1-g(x)\sim x$ is found~\cite{Bray94,Liu93}, namely a Porod's
tail $\hat g(u)\sim u^{-(d+1)}$ in momentum space for
large $u=kL(t)$. 
With Eq.~(\ref{gdirscal}) the characteristic length can then be evaluated, for instance,
as the half-height with of $G(r,t)$, namely from the condition
\be
G^{ag}(r=L(t),t)=\frac{1}{2}G^{ag}(0,t)
\label{halfh}
\ee

On usual lattices the size of domains can be easily related to the
density of interfaces $\rho (t)$. 
In fact, since domains are compact, the ratio between their
surface and their volume is proportional to $L(t)^{-1}$
and one has
$\rho(t)\sim L(t)^{-1}$. 
Since asymptotically $\rho (t)$ has
a power law behavior
\be
\rho (t)\sim t^{-\theta },
\label{rhoscal}
\ee
this implies 
\be
\theta=1/z.
\label{deltascal}
\ee
In homogeneous structures, then, $\rho (t)$ provides an indirect, 
alternative method for the determination of $L(t)$, and hence of $z$.

On generic graphs
the notion of a distance is not as straightforward as on
regular lattices and, considering the correlation function~(\ref{gdir}) 
one should, in principle, retain the full dependence
of $G_{i,j}(t)$ on $i$ and $j$. This would be a formidable task without
probably providing much insight into the physics and, in particular,
into the scaling behavior. 
Therefore, in the
following, for some of the structures considered in Sec.~\ref{structures}, 
we will attempt a {\it reasonable}
definition of a distance, we will check the validity of Eq.~(\ref{gdirscal})
with respect to this definition and we will compute $L(t)$ through Eq.~(\ref{halfh}).  
Let us also notice that the relation~(\ref{deltascal}) does
not hold on generic networks, as we will show explicitly
in our simulations. Actually, as will be discussed in Sec.~\ref{nophasetr},
the number of spins on the surface of a domain of size $L(t)$
may depend dramatically on (among other parameters) the quench temperature.
Therefore, in general, there is not a unique relation between
$\rho (t)$ and $L(t)$ and they provide independent informations.

The two time quantities that will be considered in this paper
are the (spatially averaged) autocorrelation function
\be
C(t,s)={1\over N} \sum_{i=1}^N \langle  \sigma_i (t)\cdot \sigma_i (s) \rangle
\label{autocorr}
\ee
and the integrated (auto)response function  
\be
\chi (t,s)=\int _s ^t dt' R(t,t').
\label{integrated}
\ee
The quantity 
\be
R(t,t')= {1\over N} \sum_{i=1}^N \left . \frac {\partial \langle \sigma_i (t) \rangle}
{\partial h_i (t')}\right \vert _{h=0},
\ee
is the (spatially averaged) linear response function associated 
to the perturbation caused by an impulsive magnetic field $h_i$
switched on at time $t'<t$.

On regular lattices, scalings~(\ref{2},\ref{2a}) imply
\be
C^{ag}(t,s)=h(y)
\label{scalauto}
\ee
and
\be
\chi^{ag}(t,s)=s^{-a_\chi }f(y),
\label{scalresp}
\ee
where $y=t/s$ and $h(y)=\tilde {\cal C}(0,y)$, $f(y)=\tilde \chi(0,y)$,
and the large-$y$ behaviors
\be
h(y)\sim y^{-\lambda },
\label{largeyc}
\ee
\be
f(y)\sim y^{-a_\chi }
\label{largeychi}
\ee

In order to compute $\chi (t,s)$ we enforce
the out of equilibrium generalization
of the fluctuation dissipation theorem derived in~\cite{Lippiello05}, 
which relates the response function
to particular correlation functions of the unperturbed system
\be
T\chi (t,s)=\frac {1}{2}\left [C(t,t)-C(t,s)-
\int _s ^t \langle \sigma_i(t)B_i(t')\rangle dt' \right ].
\label{algochi}
\ee
where
\be 
B_i[\sigma]=-\sum _{\sigma '}(\sigma_i-\sigma_i') w([\sigma]\to [\sigma']).
\label{b}
\ee
In this equation $[\sigma]$ and $[\sigma']$ are two configurations differing only by
the spin on site $i$, taking the values $\sigma_i$ and $\sigma_i'$ respectively.   
This relation allows to compute the integrated response function by
measuring correlation functions on the unperturbed system, avoiding the
complications of the traditional methods where a perturbation is
applied, and improving significantly the quality of the 
results~\cite{Lippiello05}. 

\section{Numerical simulations} \label{structures}

In the following we will present the numerical results. 
We set $J=1$.
Statistical errors are comparable to the thickness of the symbols. We recall that NBF dynamics 
is always used, and therefore we measure directly the aging part of
every observable considered.

\subsection{Graphs with $T_c=0$} \label{nophasetr}

Let start with systems of class II, first.
The structures considered in the simulations will be the Sierpinski gasket (SG),
the T-fractal (TF) and the percolation cluster (PC), namely a diluted square 
lattice, at the the percolation threshold (see Fig.~\ref{strutture}). 
The number $N$ of spins in this structures
is 265722 and 531442 for the SG, and TS, respectively. The PC has been obtained by means of site percolation on a square lattice of size $1200\time 1200$ at the critical dilution $p_c=0.407$
 The results presented are 
thermal averages over $100$ realizations.

For the SG we have computed the equal 
time correlation function~(\ref{gdir}) by restricting
$i$ and $j$ along the borders of the structure. This allows a natural definition
of the distance $r$ between $i$ and $j$, since along this lines all sites are
occupied by spins. 
For the TF we have computed $G^{ag}(r,t)=G_{ij}^{ag}(t)$ for points $i$ and $j$ along the 
the baseline of the structure.

\begin{figure}
    \centering
   \rotatebox{0}{\resizebox{0.8\textwidth}{!}{\includegraphics{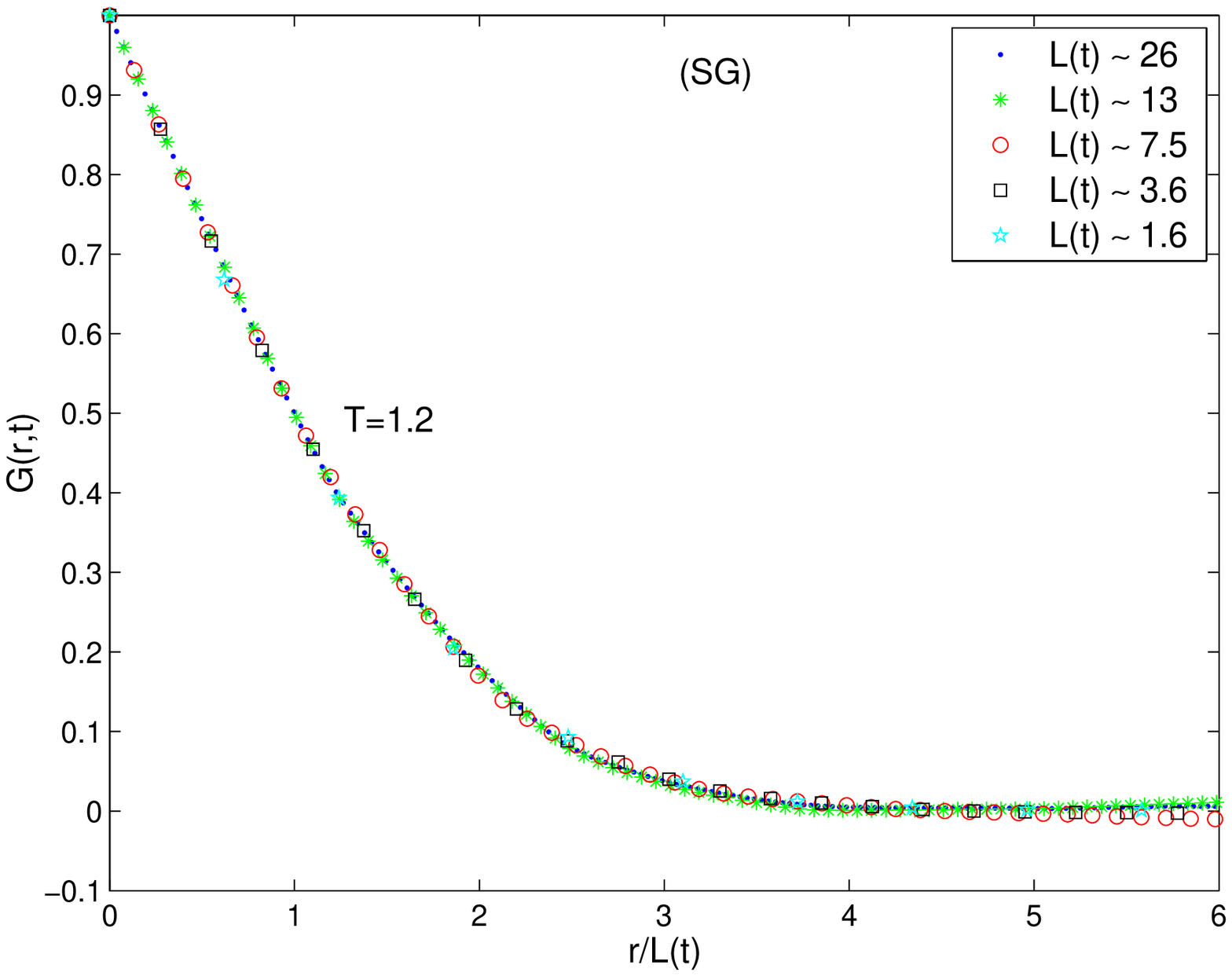}\includegraphics{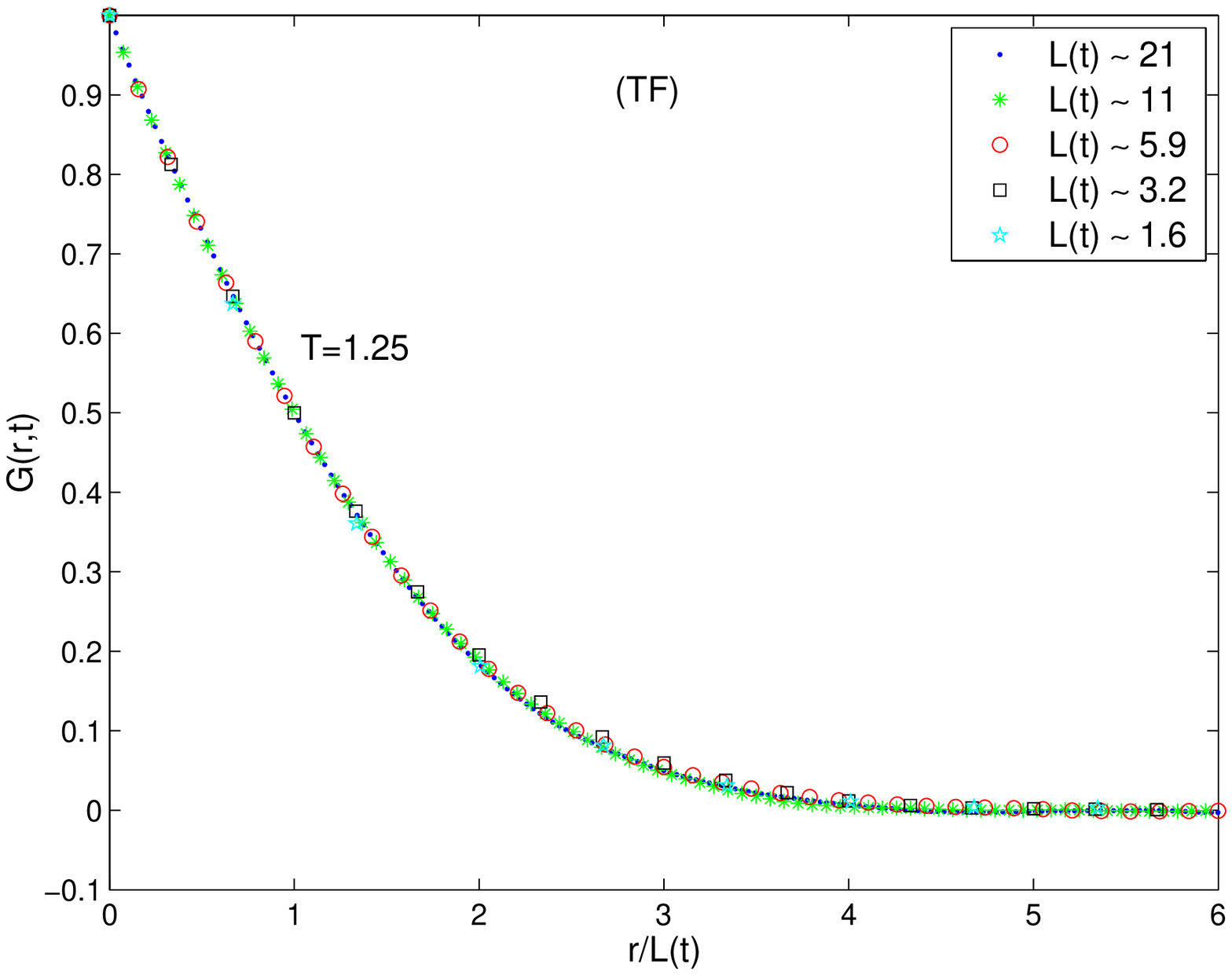}}}
    \caption{$G^{ag}(r,t)$ is plotted against $x=r/L(t)$ for the (TF) and  the (SG), the time $t$ ranges from $10$ steps for the shortest $L(t)$ to $400000$ for the longest lengths.}
\label{scalg1}
\end{figure}

\begin{figure}
    \centering
    \rotatebox{0}{\resizebox{0.8\textwidth}{!}{\includegraphics{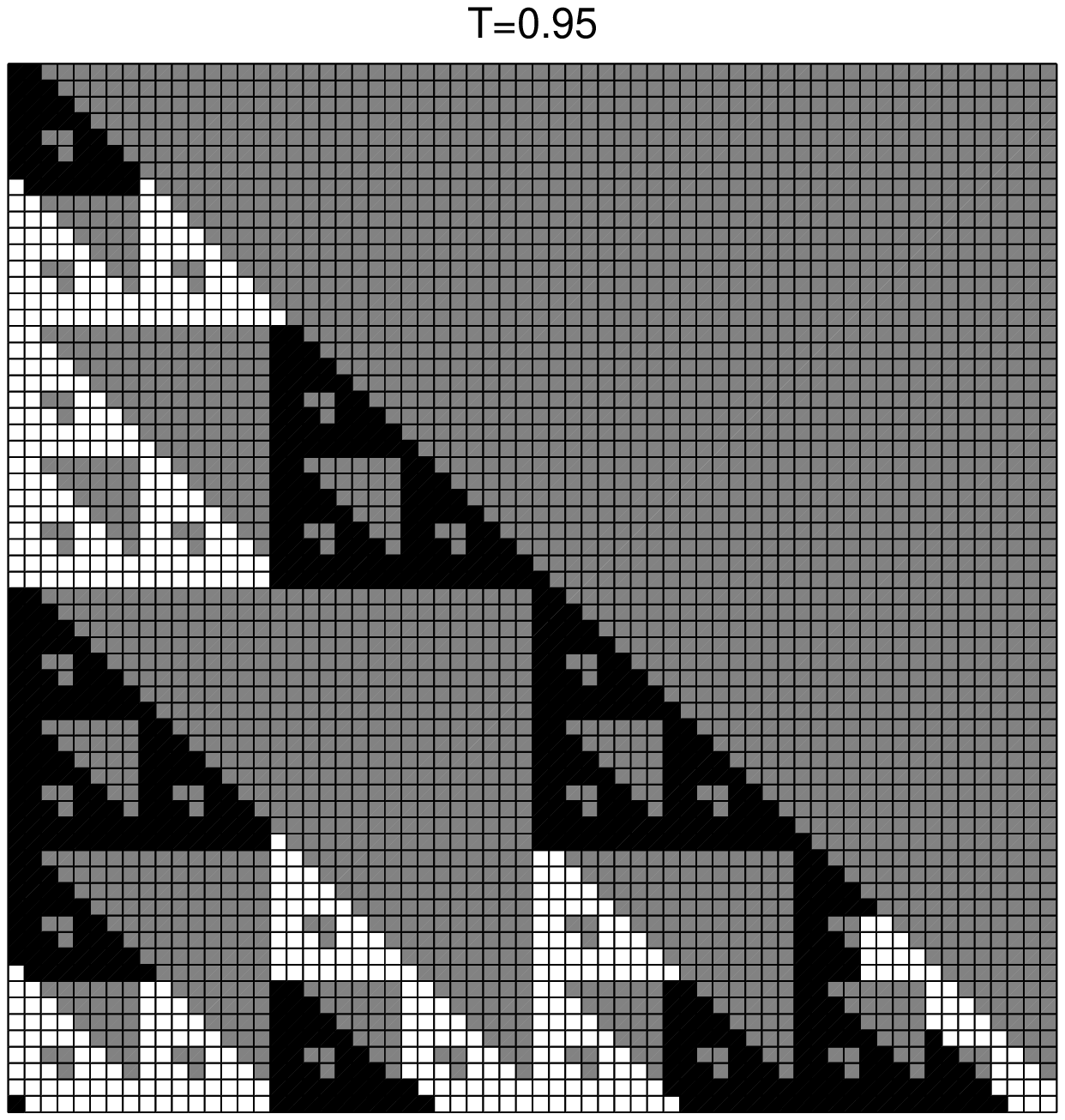}\includegraphics{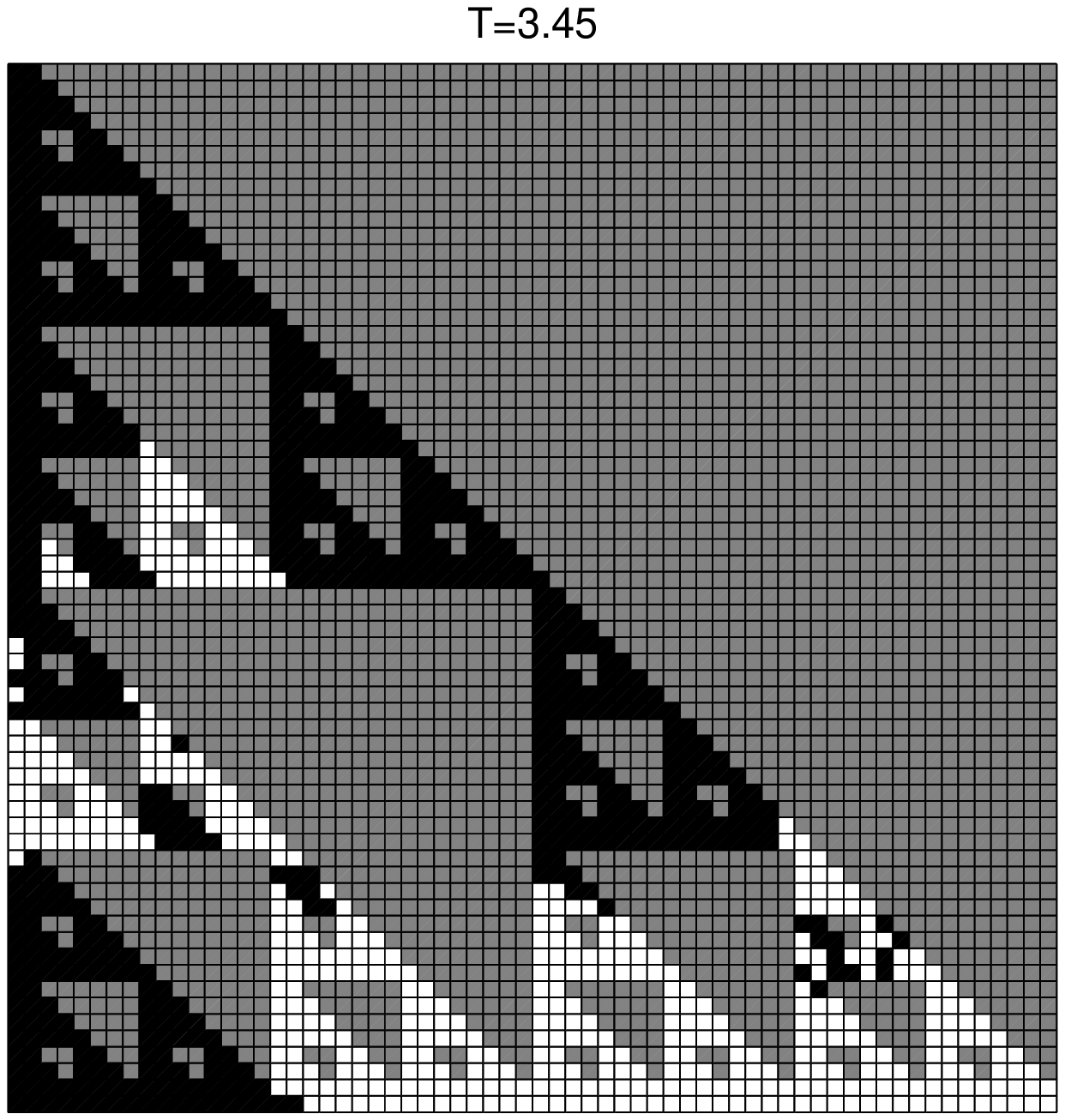}}}
    \caption{(SG) Two spin configurations after $t=1000$ step for the temperatures $T_f=0.95$ and $T_f=3.45$.}
\label{intergas}
\end{figure}

\begin{figure}
    \centering
   \rotatebox{0}{\resizebox{.4\textwidth}{!}{\includegraphics{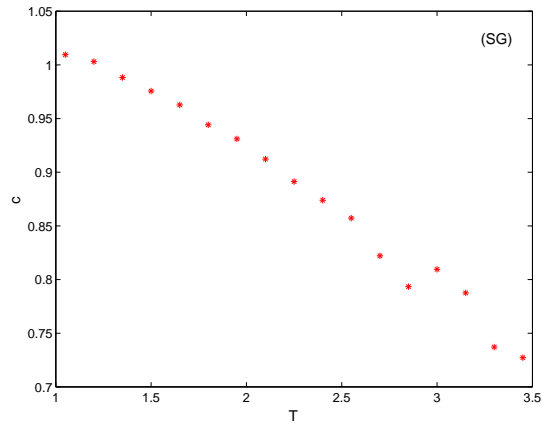}}}
    \caption{(SG) The exponent $c$ characterizing the behaviour of $g(x)$
    for $x\simeq 0$.}
\label{expporodgasc}
\end{figure}

In Fig.~\ref{scalg1}, $G^{ag}(r,t)$ is plotted against
$x=r/L(t)$ for the SG and the TF. One finds a very good data collapse, as expected on the basis
of Eq.~(\ref{gdirscal}). This indicates that dynamical scaling is obeyed
also on these structures. Notice also the presence of the Porod's tail, for small
$x$, implying that interfaces are sharp objects at the relatively low temperatures considered
in Fig.~\ref{scalg1}. However, when the temperature is raised, 
the interfaces broaden on the fractal substrate, as shown in Fig. \ref{intergas}. 
Here one expects a 
deviation from the Porod linear behavior. This can be clearly seen in Fig. \ref{expporodgasc}.
Here we plot the exponent $c$ which regulates the small $x$ decay of $g(x)$,
namely $1-g(x)\sim x^c$, as a function of $T_f$.
Clearly, given the choice of $i$,$j$ described above, our considerations strictly apply
only along the particular boundary of the structure where $G^{ag}(r,t)$ is computed. 
However, we expect this results to be representative of the whole system, due to the scale
invariance of the structure.

\begin{figure}
    \centering
   \rotatebox{0}{\resizebox{0.8\textwidth}{!}{\includegraphics{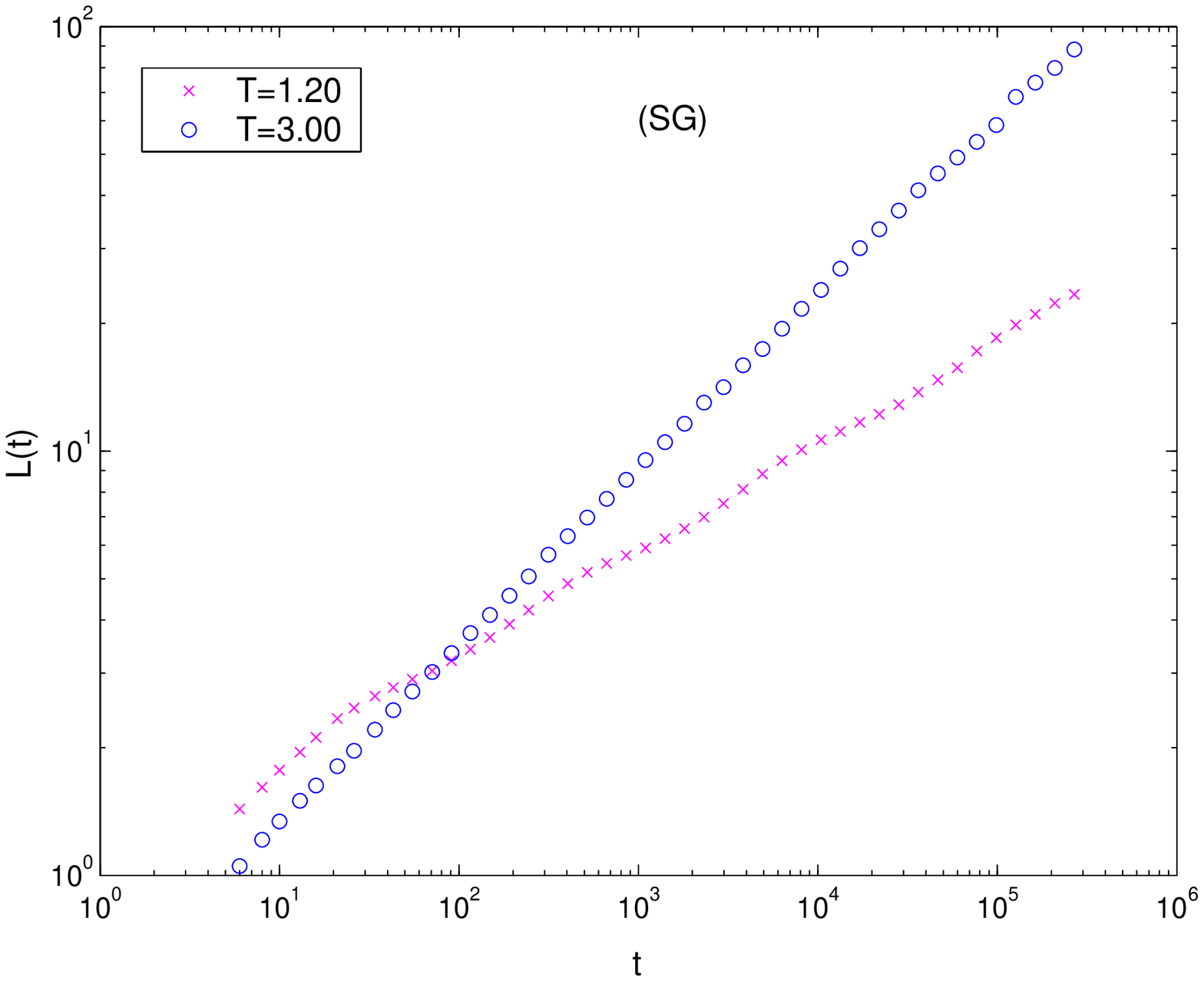}\includegraphics{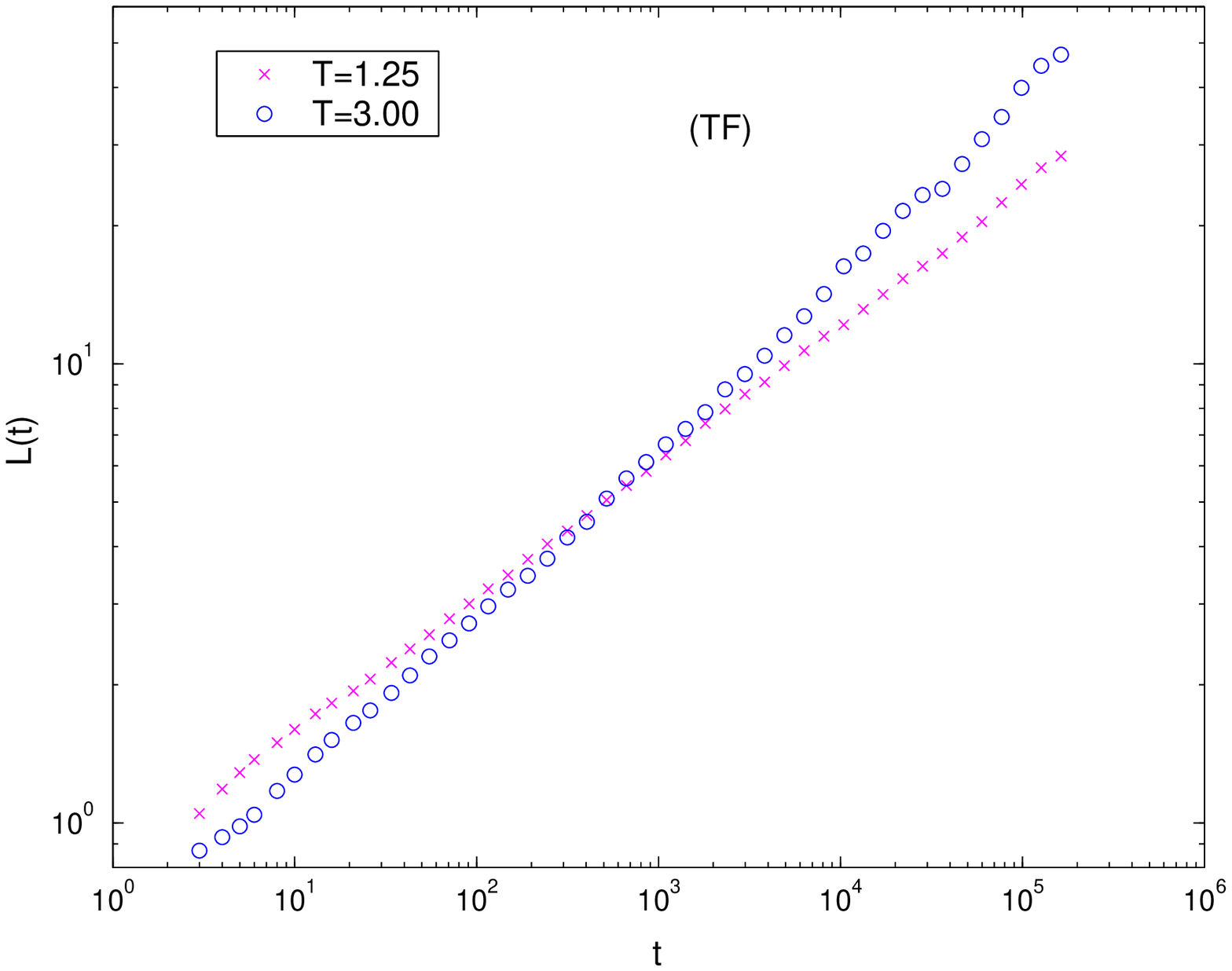}}}
    \caption{ The characteristic length $L(t)$ is plotted against $t$, for two
    different temperatures $T_f=1.2$ and $T_f=3.0$ for the (SC) and $T_f=1.25$ and $T_f=3.0$ for the (TF).}
\label{length1}
\end{figure}

\begin{figure}
    \centering
   \rotatebox{0}{\resizebox{.4\textwidth}{!}{\includegraphics{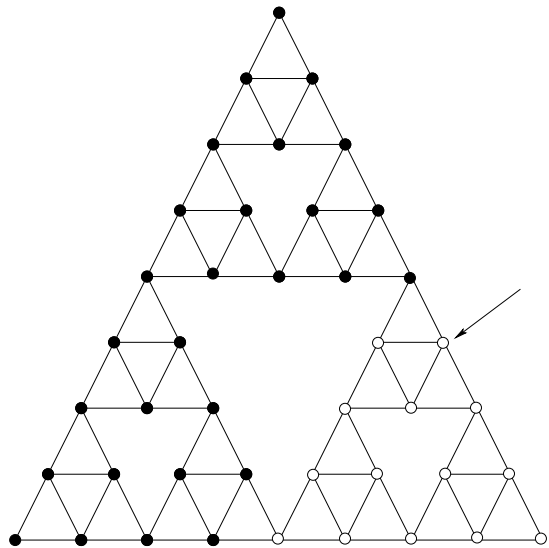}}}
    \caption{Schematic representation of a part of the structure, showing
             the necessity of activated processes.}
\label{schemagas}
\end{figure}

The characteristic size $L(t)$, obtained as in Eq.~(\ref{halfh}), is
shown in Fig.~\ref{length1}. This length grows roughly as a power law, but with 
a superimposed oscillation which is more pronounced at low temperatures.
The origin of such oscillation, which is very reminiscent of what one
observes in the related problem of Brownian motion on this structure \cite{woess},
are again the pinning forces. In fact, as shown schematically in Fig.~\ref{schemagas}
for the SG,
if two triangles happen to be ordered differently, in order to start
reversing one of the two to achieve a global ordering, one has to flip
one of the interfacial spins, for example the one marked with an arrow. This move requires an activation
energy $\Delta E=4J$ and can then be accomplished on a time 
$\tau \simeq \exp (4J/T_f)$. The dynamics then proceeds without activated
processes until all the triangle is reversed but, then, a new activated
step is required and so on. During the time $\tau $ the
dynamics on the triangle in consideration is frozen. Since all the triangles
have the same qualitative behavior, with some fluctuations, 
the overall behavior of $L(t)$ shows a periodic oscillation due to the recurring
slow down caused by activated processes. 
An analogous phenomenon occurs in the TF.
Clearly, since $\tau$
decreases at larger temperatures, this phenomenon is less evident, although
still present even at relatively high temperatures, as shown on 
Fig.~\ref{length1}.

\begin{figure}
    \centering
   \rotatebox{0}{\resizebox{0.8\textwidth}{!}{\includegraphics{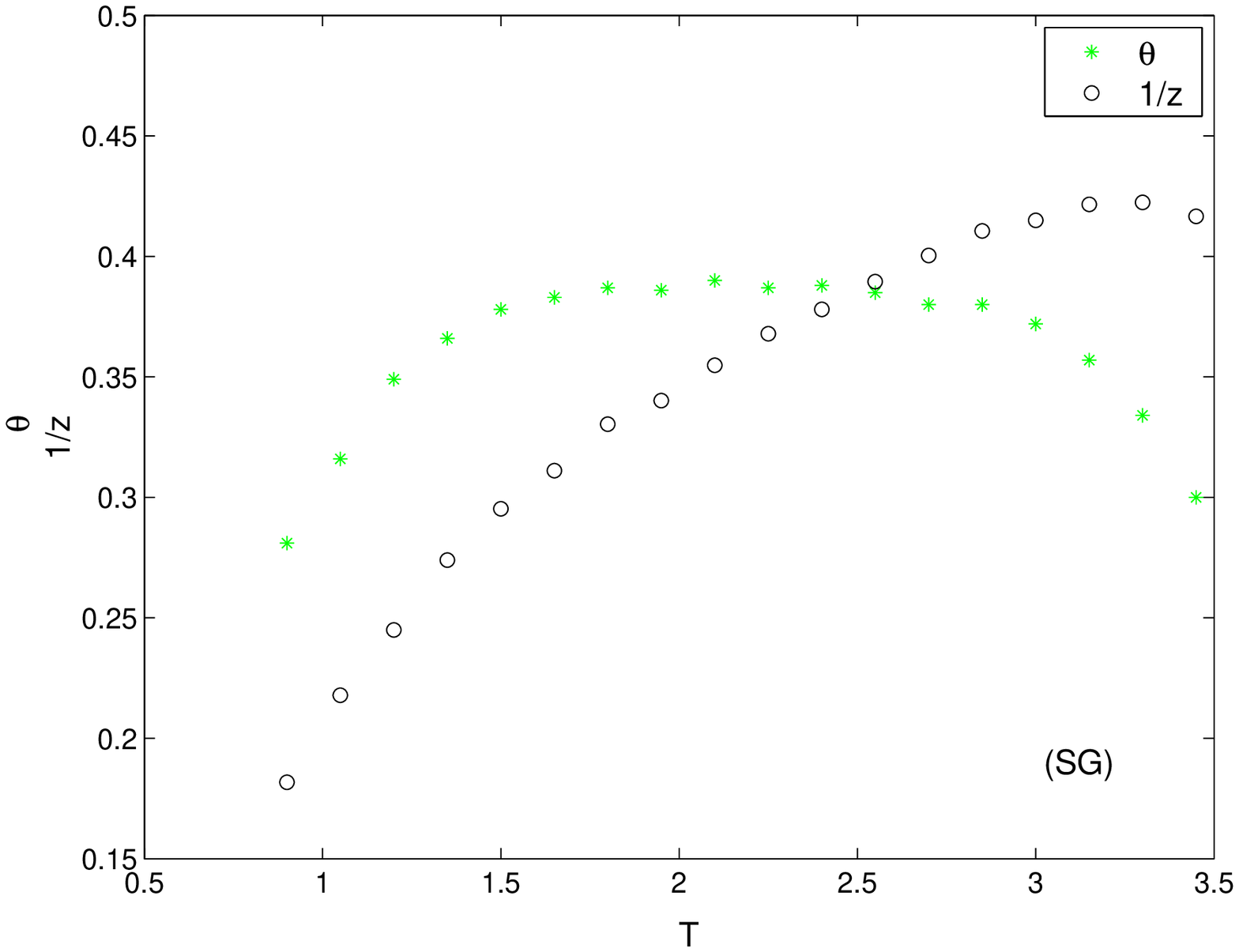}\includegraphics{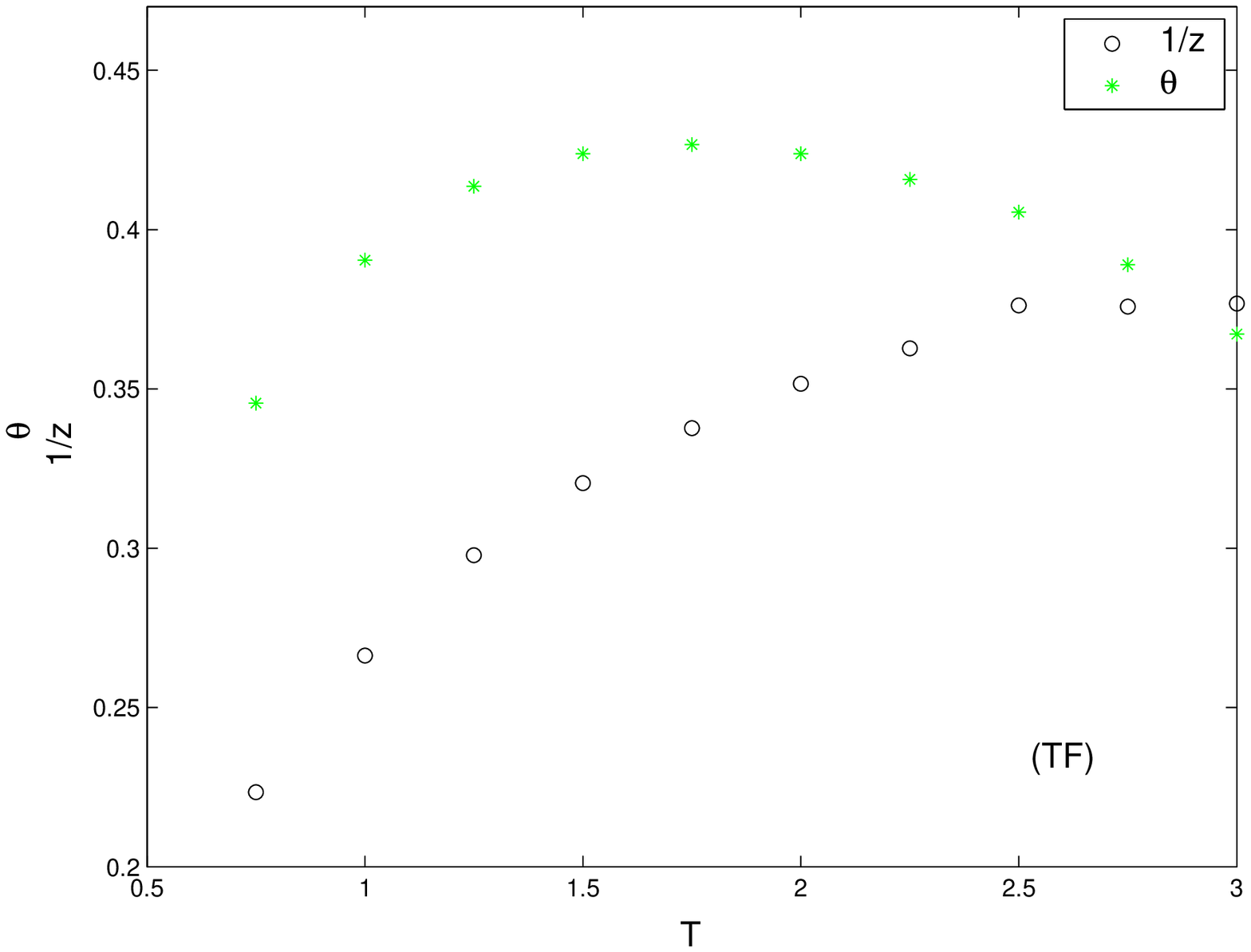}}}
    \caption{The dependence on temperature of the exponents $1/z$ and $\theta$ for the (SG) and the (TF).}
\label{exp}
\end{figure}

The curves of Fig.~\ref{length1} can be quite convincingly
fitted by a power law  with periodic oscillations superimposed.
The exponent $z$ of the power law, however, should be thought as an
effective exponent resulting from the balance between non activated
processes obeying a growth law~(\ref{ldit}) with a certain value of $z$, 
and the activated processes whose net effect is an overall slowing down. Since the relative
importance of these two processes is regulated by $T_f$, $z$ turns out to
be temperature dependent, as shown in Fig.~\ref{exp}. On the basis 
of the previous discussion, one would expect to have a faster growth,
namely a larger $1/z$, for higher $T_f$. This is indeed observed in Fig.~\ref{exp}
in a broad range. For very large temperatures $1/z$ seems to saturate. 
This is perhaps due to the neighborhood of the critical
temperature (that with NBF turns out to be $T_c\simeq 3.6$ and $T_c\simeq 2$ for the
SG and the TF, respectively),
slowing down the dynamics much in the same way as it happens on regular lattices.

\begin{figure}
    \centering
   \rotatebox{0}{\resizebox{.8\textwidth}{!}{\includegraphics{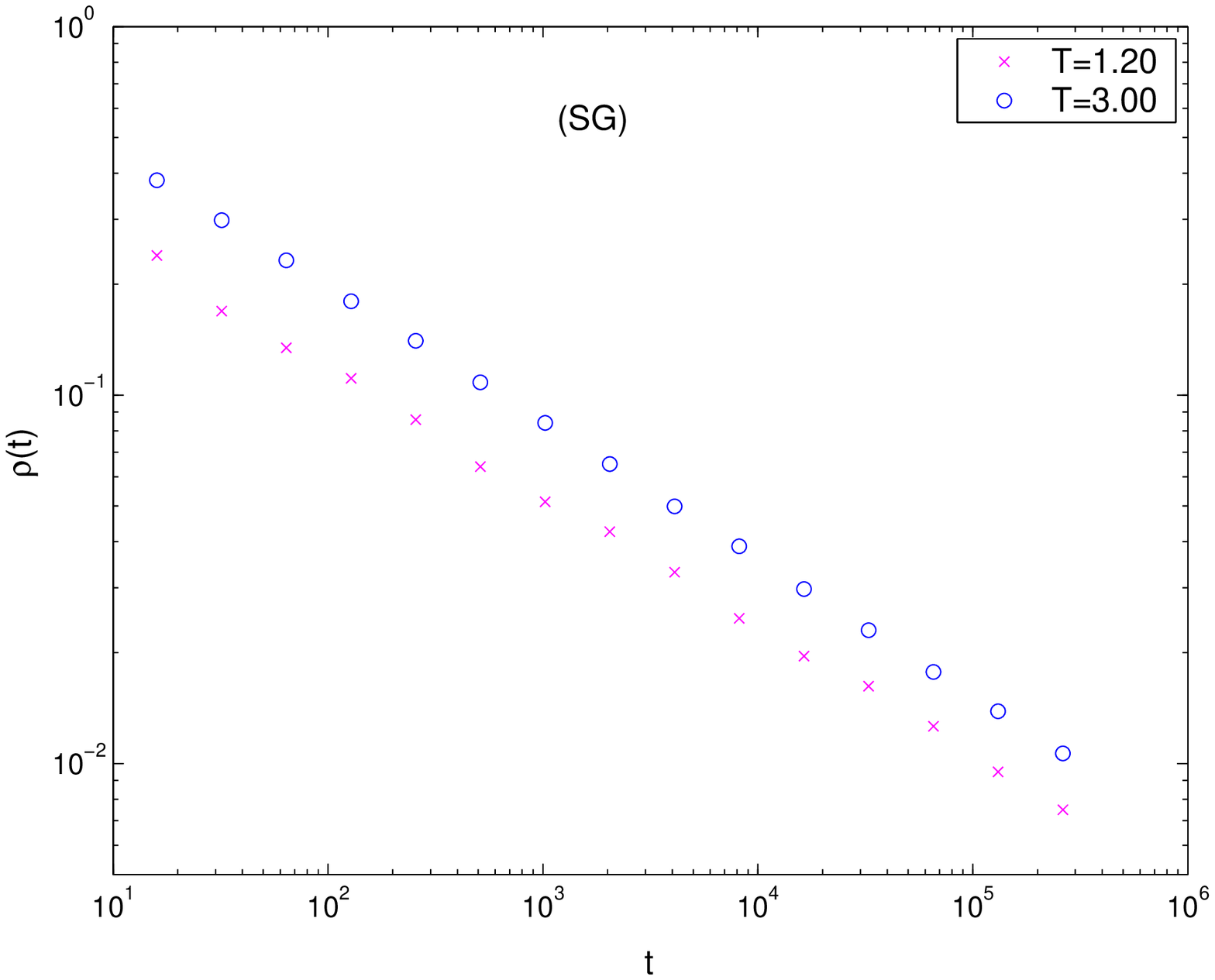}\includegraphics{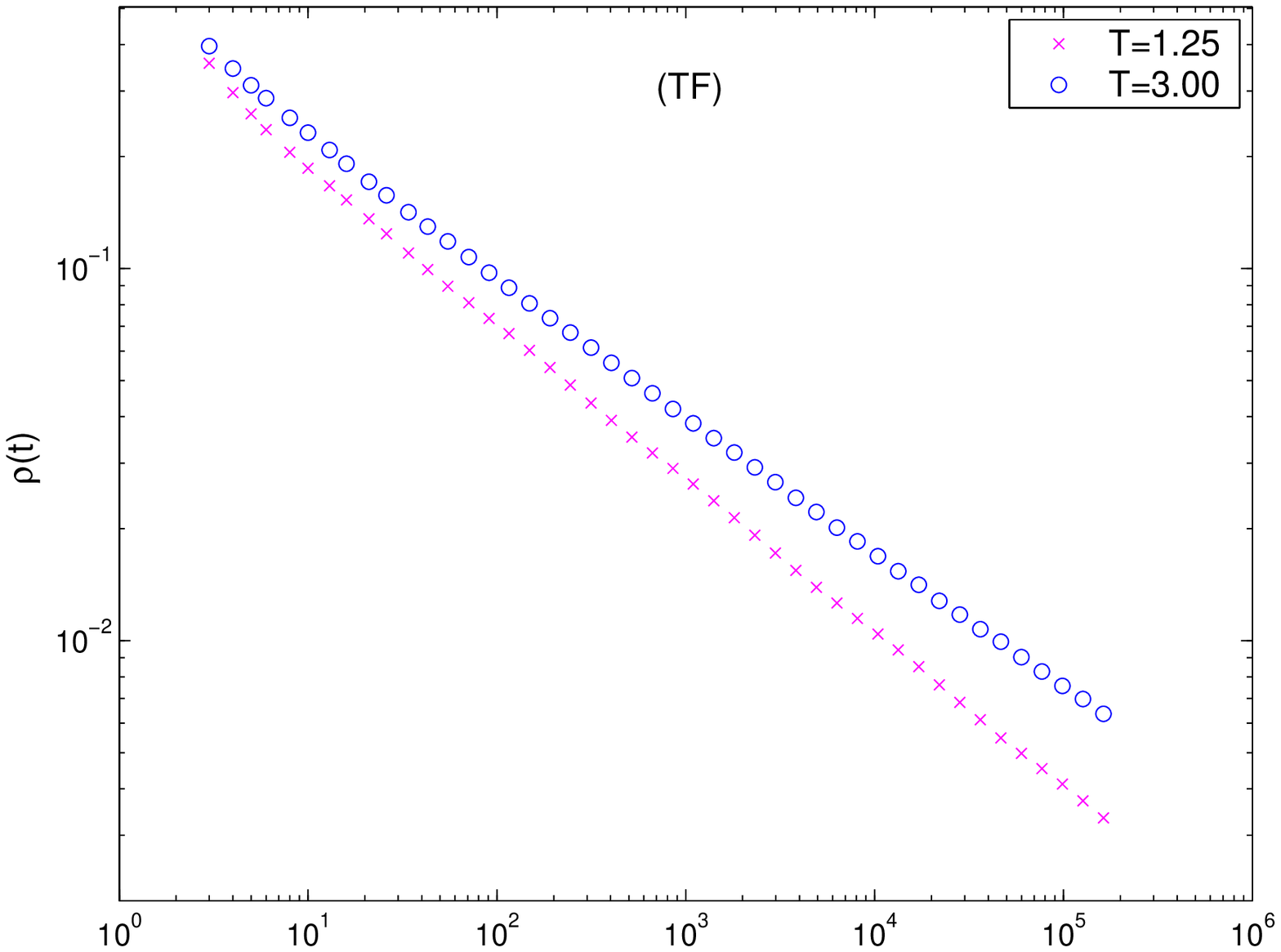}}}
 \rotatebox{0}{\resizebox{.4\textwidth}{!}{\includegraphics{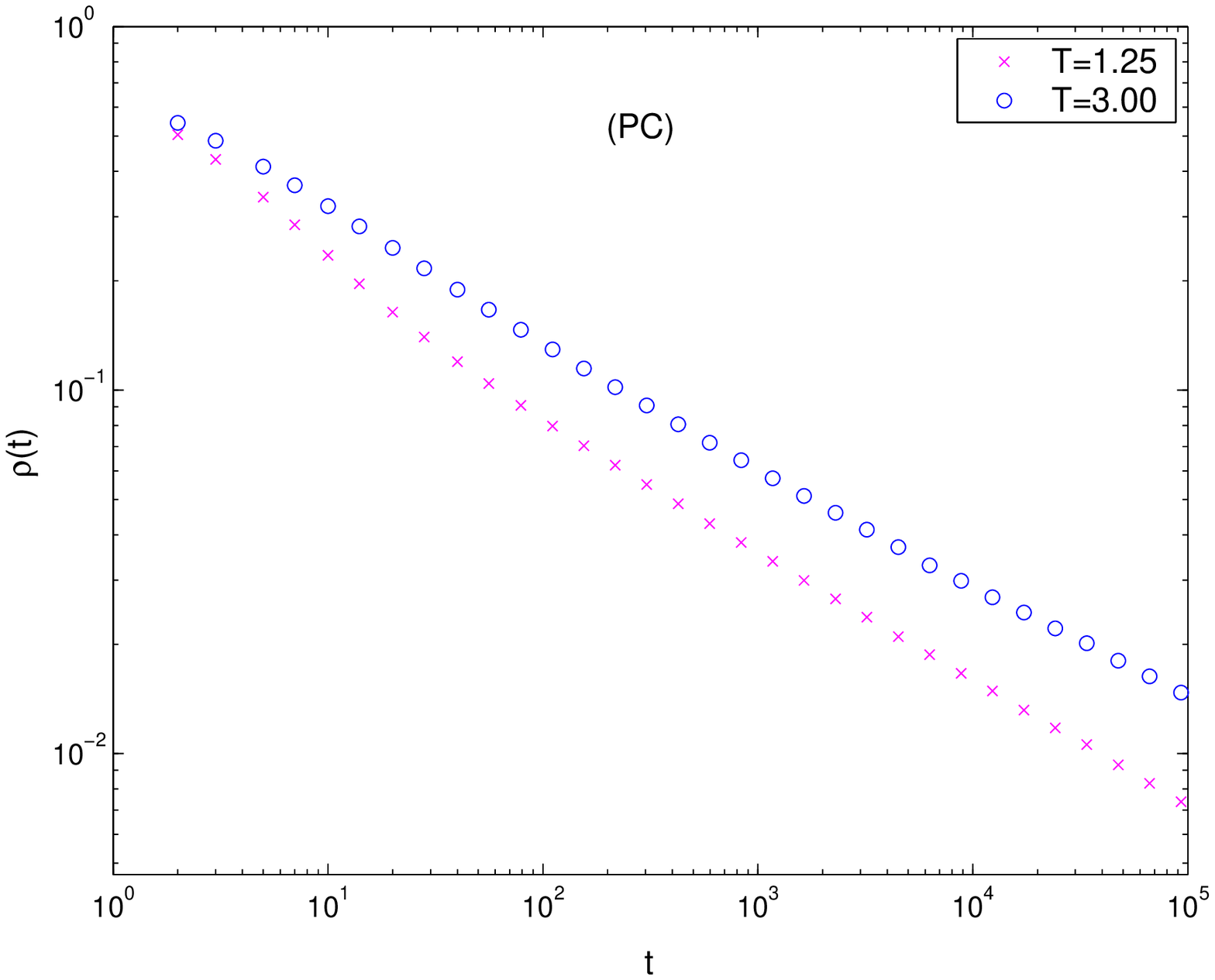}}}
    \caption{The density of interfaces $\rho (t)$ is plotted against $t$, at the temperatures $T_f=1.2$ and $T_f=3.0$ for the (SG),  at $T_f=1.25$ and $T_f=3.0$ for the (TF) and (PC).}
\label{bordi1}
\end{figure}

A pattern similar to that of $L(t)$ is observed for $\rho (t)$. This quantity,
defined as the density of spin which are not fully aligned with their neighbors,
is shown in Fig.~\ref{bordi1}  for the SG, TF and PC. Also in this case one observes oscillations on top of the power law behavior~(\ref{rhoscal}).
The value of $\theta$ also depends on temperature,
as shown in Fig.~\ref{exp}. Differently from $1/z$, however, the behavior is
strongly non monotonous, with a broad maximum at intermediate temperatures.
As already observed, the exponent $\theta $ and $1/z$ are not trivially related,
as on homogeneous structures. In the limit of low temperatures
one finds an exponent $\theta $ consistent with $\theta = -d_f/z$, which
can be understood on the basis of the following argument. Let us consider
again, for simplicity, the SG of Fig.~\ref{schemagas}.
When the temperature is very low interfaces are very likely located in the
pinning positions, as can be seen in Fig.~\ref{intergas}. A domain, therefore, is a triangle of size $L(t)$, with a 
volume $V\propto L(t)^{d_f}$ and a number of surface spins $n_s=3$. The density
of interfacial spins is, therefore $\rho (t)\sim L(t)^{-d_f}$. One then has
\be
\theta = -\frac{d_f}{z}
\label{deltascal1},
\ee
instead of Eq.~(\ref{deltascal}), holding on regular lattices.
Let us remark again, however, that, although this relation is consistent with
our data in the limit of small $T_f$, it is not of general validity.
Actually, as already evident from Fig.~\ref{intergas} and from the non linear 
behavior of $g(x)$ at small $x$, 
for larger temperatures, interfaces are no longer located on the pinning centers. 
\begin{figure}
    \centering
   \rotatebox{0}{\resizebox{.8\textwidth}{!}{\includegraphics{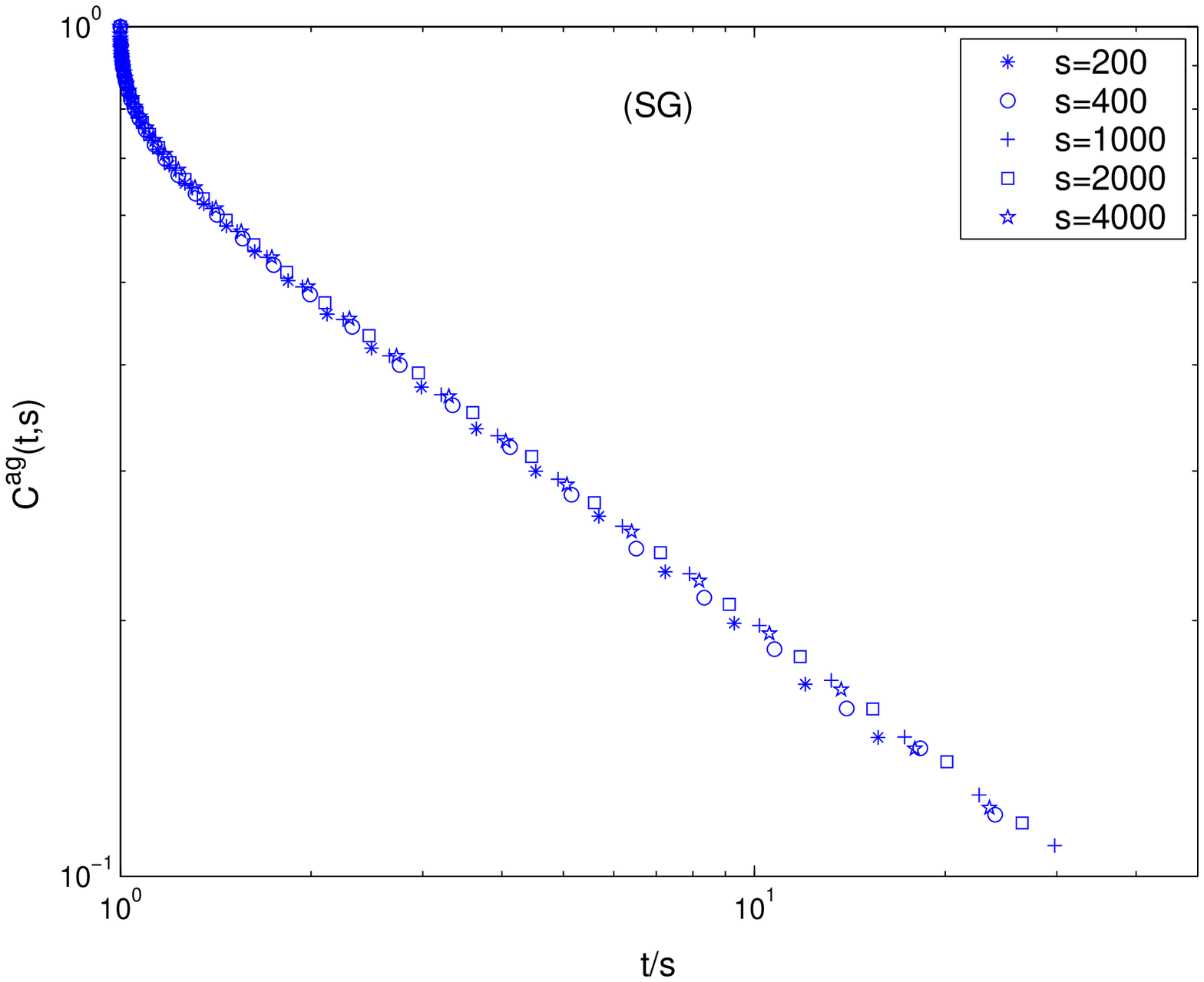}\includegraphics{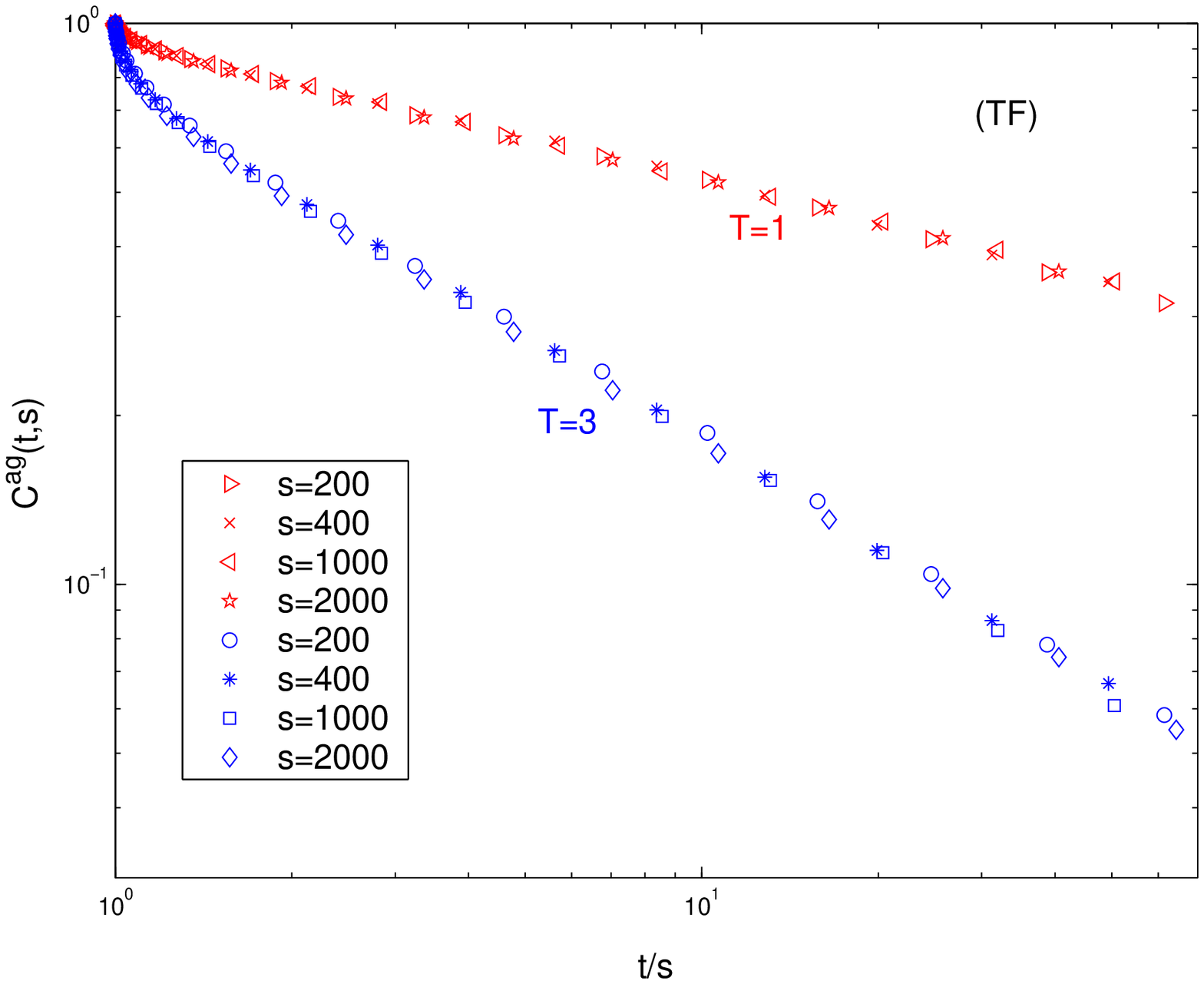}}}
\rotatebox{0}{\resizebox{.4\textwidth}{!}{\includegraphics{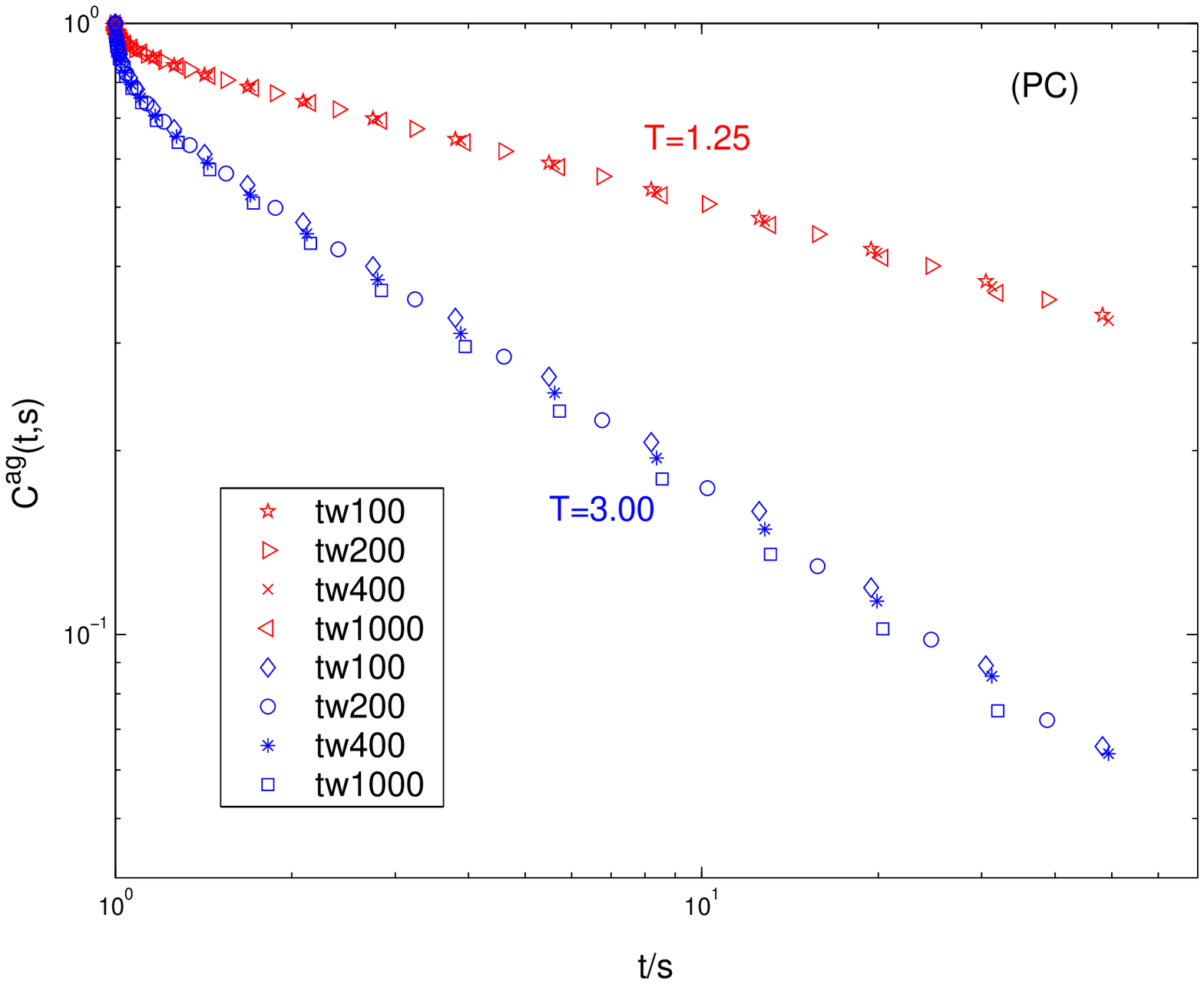}}}
     \caption{(SG) $C^{ag}(t,s)$ is plotted against $y=t/s$, at 
    $T_f=3.0$ for the (SG), at $T_f=3.0$ and $T_f=1.0$ for the (TF) and 
    at $T_f=3.00$ and $T_f=1.25$ for the (PC).}
\label{scalc1}
\end{figure}

Let us consider now the behavior of two time quantities.
The autocorrelation function for the SG, TF and PC are plotted in 
Fig.~\ref{scalc1} against $y=t/s$. According to Eq.~(\ref{scalauto}), one should observe data collapse of the curves
with different $s$. This is indeed observed.
One also finds the
large-$y$ power law behavior~(\ref{largeyc}) with an exponent $\lambda $ strongly
dependent on temperature. 

\begin{figure}
    \centering
   \rotatebox{0}{\resizebox{.4\textwidth}{!}{\includegraphics{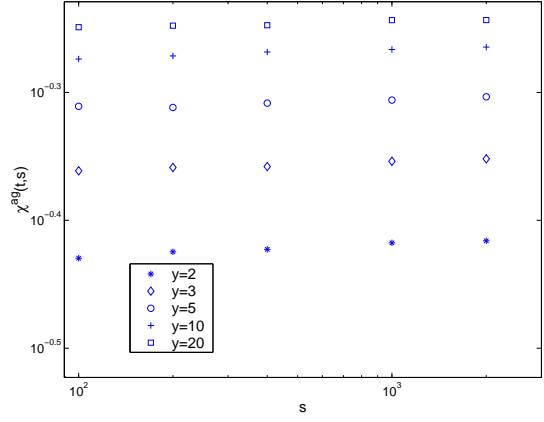}}}
    \caption{(TF) $\chi ^{ag}(t,s)$ is plotted against $s$ for fixed values of $y=t/s$, for 
    $T_f=3.00$.}
\label{scalchi1tf}
\end{figure}

\begin{figure}
    \centering
   \rotatebox{0}{\resizebox{.8\textwidth}{!}{\includegraphics{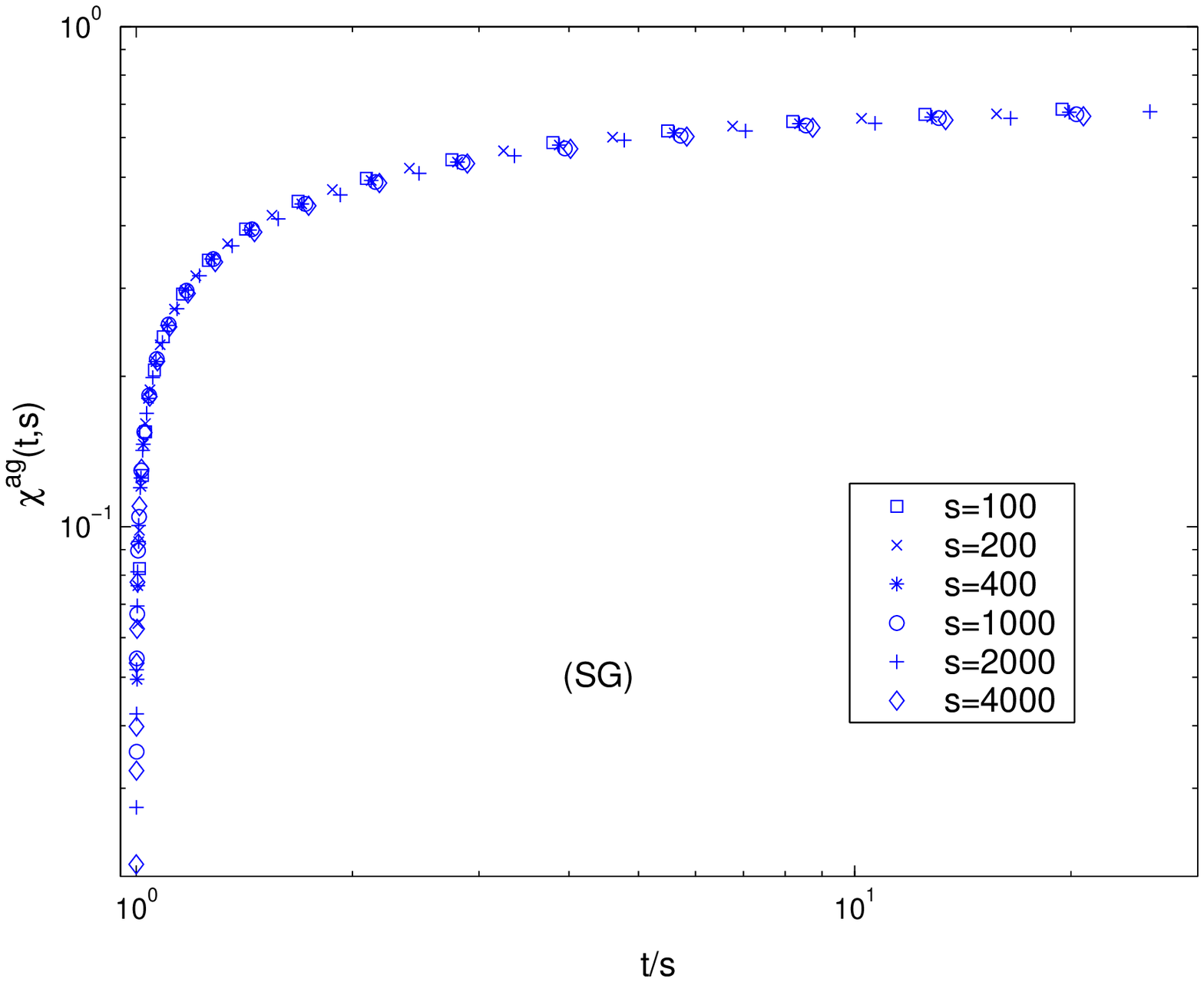}\includegraphics{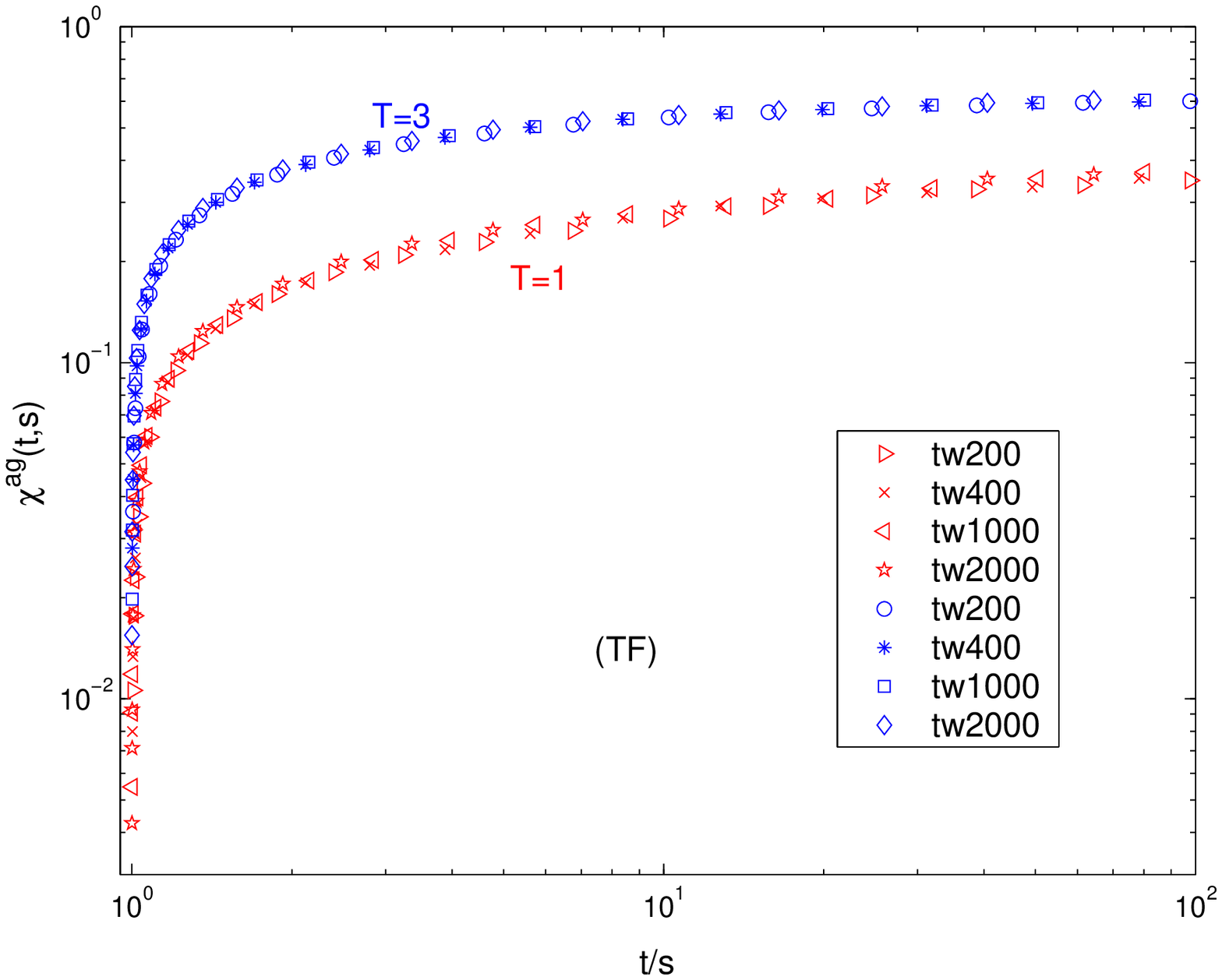}}}
\rotatebox{0}{\resizebox{.4\textwidth}{!}{\includegraphics{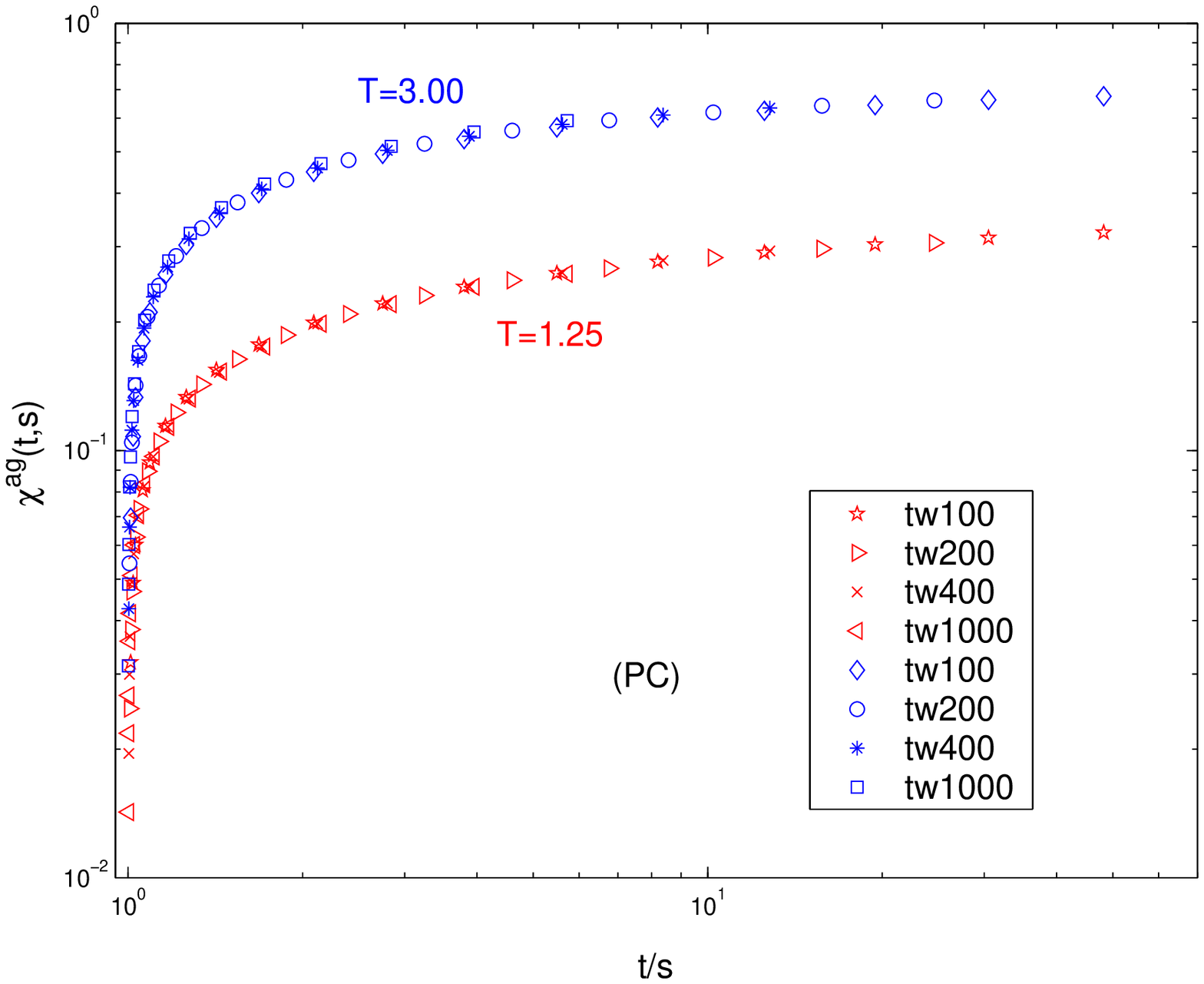}}}
    \caption{(SG) $\chi ^{ag}(t,s)$ is plotted against $y=t/s$ for different values of $s$, for 
    $T_f=3.00$ (SG),  $T_f=3.00$ and $T_f=1.00$ (TF) and for $T_f=3.00$ and $T_f=1.25$ (PC).}
\label{scalchi1}
\end{figure}

Let us turn to consider the response function.
According to Eq.~(\ref{scalresp}) the exponent $a_\chi $ can be obtained as
the slope of a double logarithmic plot of $\chi ^{ag}(t,s)$ against $s$,
with $y$ held fixed. Such determination should be independent on $y$,
within errors. We show this plot in Fig.~\ref{scalchi1tf} for the TF.
We obtain $a_\chi = .01 \pm .02 $ which is
consistent with $a_\chi =0$, similarly to what found in homogeneous 
systems with $T_c=0$~\cite{Lippiello00,Godreche00,Castellano04}. Analogous values are found for
all the structures with $T_c=0$. Then, by plotting $\chi ^{ag}(t,s)$ against $y$ one should find data collapse,
how it is indeed shown in Fig.~\ref{scalchi1} for SG, TF and PC at different temperatures. 

\subsection{Graphs with $T_c>0$} \label{phasetr}

In this Section we will consider systems of class I. In particular we will
study the phase-ordering kinetics on 
the diluted square lattice (DS) above the percolation threshold, the
{\it toblerone} lattice (TL) obtained by replicating the Sierpinski
gasket along a third spatial direction $z$,
and the Sierpinski carpet (SC). 
These structures are represented 
in Fig.~\ref{strutture}.

\begin{figure}
    \centering
    \rotatebox{0}{\resizebox{0.8\textwidth}{!}{\includegraphics{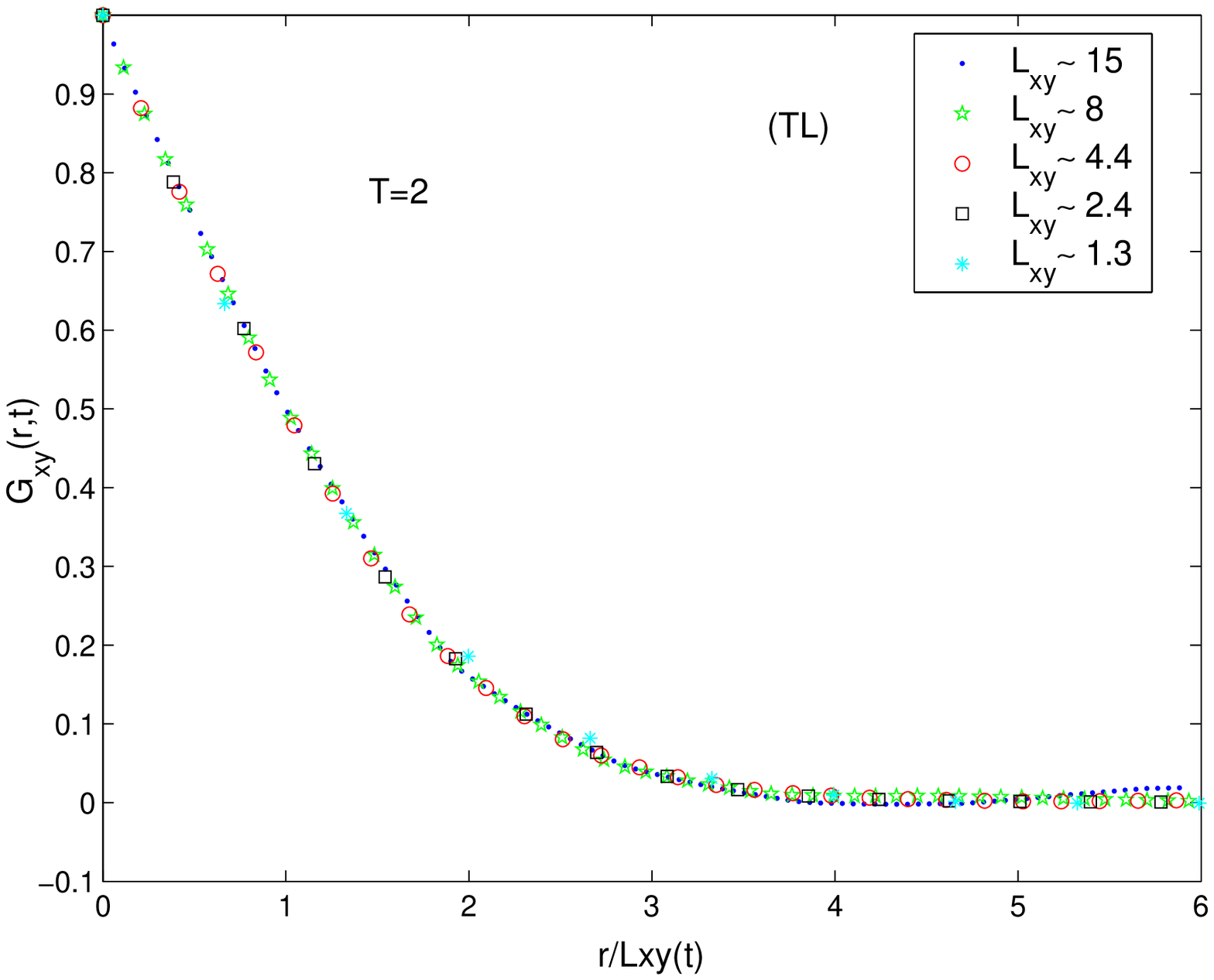}\includegraphics{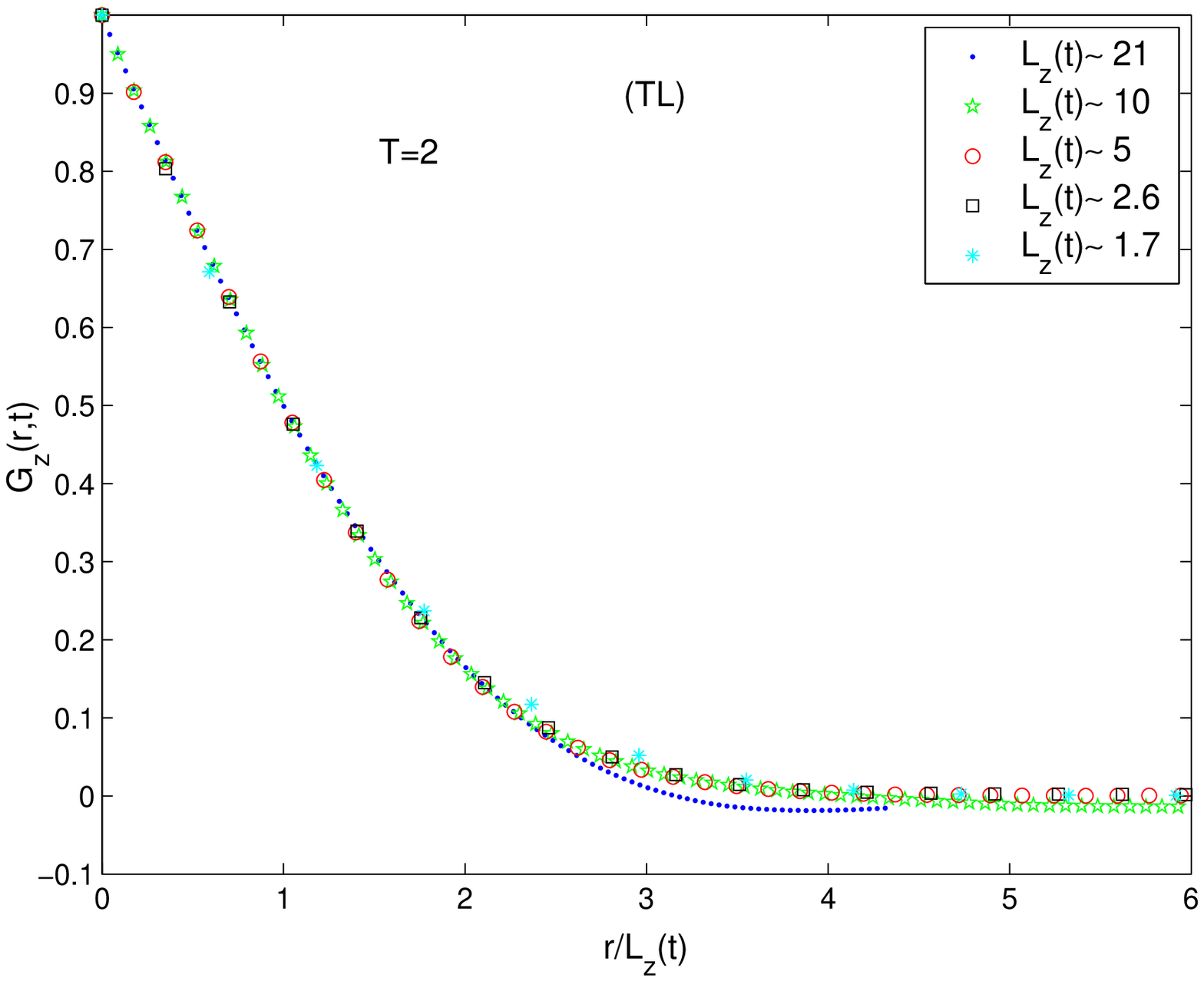}}}
\rotatebox{0}{\resizebox{.4\textwidth}{!}{\includegraphics{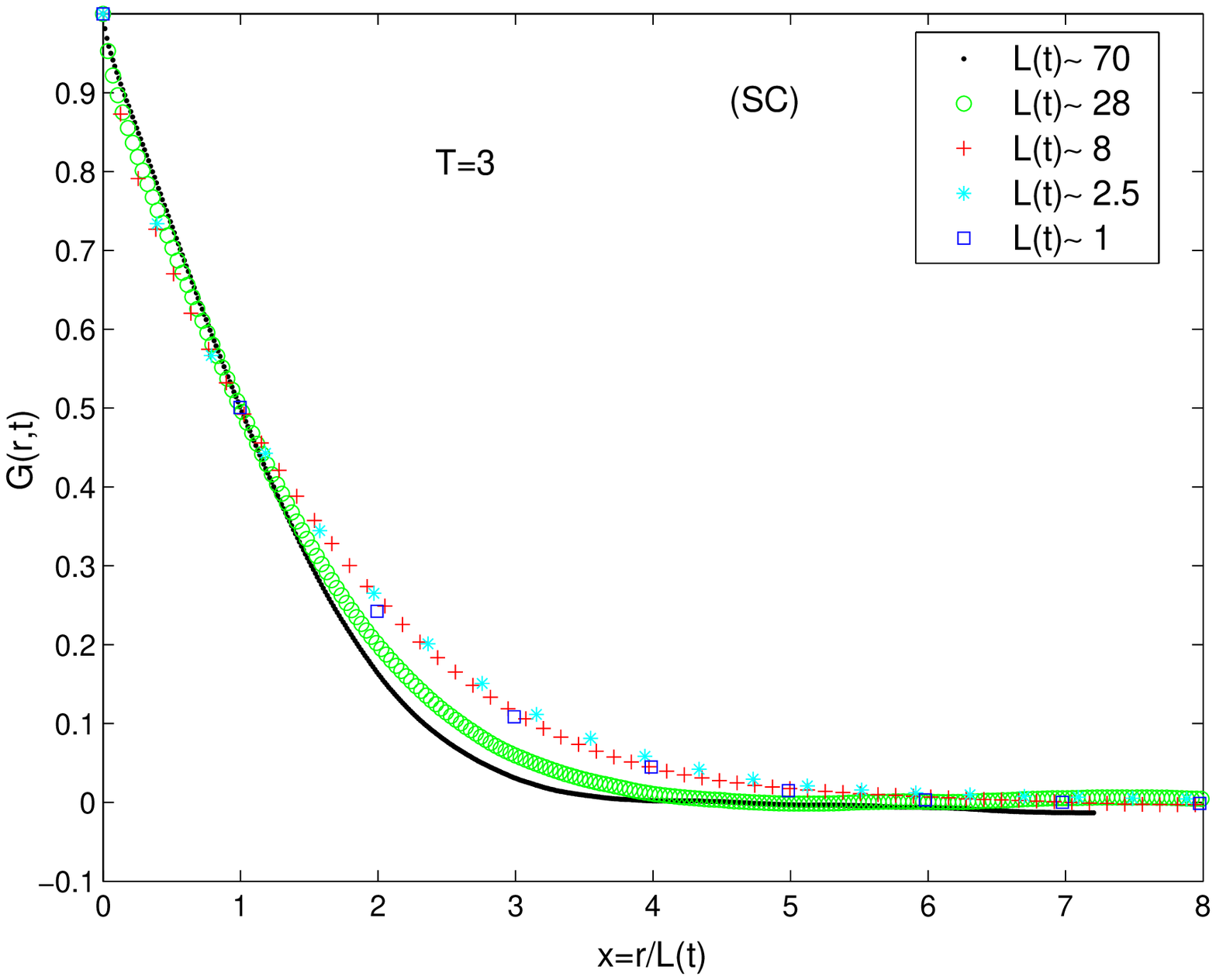}}}
    \caption{For the (TL) we plotted $G^{ag}_{xy}(r,t)$ and $G^{ag}_{z}(r,t)$ 
against $x=r/L(t)$ for the (SC) we plotted $G^{ag}(r,t)$.}
\label{scalg2}
\end{figure}

For the SC, we have computed the equal 
time correlation function~(\ref{gdir}) 
along the borders of the structure, analogously to
what done for the SG. 
The TL is an anisotropic structure, since it is translational
invariant along the $z$ direction alone. Therefore we have computed 
the equal time correlation function on the planes with a constant
$z$, and in the $z$ direction separately. These are denoted
as $G^{ag}_{xy}(r,t)$ and $G^{ag}_{z}(r,t)$, respectively.
The former is computed along the borders of the structure,
as for the SG. In Fig.~\ref{scalg2} 
$G^{ag}(r,t)$ is plotted against
$x=r/L(t)$ for the TL and the SC. One finds a very good data collapse for the TL, 
as expected on the basis
of Eq.~(\ref{gdirscal}). The scaling is less accurate for the SG, where the pinning effects
seem to be stronger.
Notice also in these cases, the presence of the Porod's tail, for small
$x$, implying that interfaces are sharp objects at low temperature. As in the case of the
SG, when the temperature is raised the interfaces broaden and the Porod behavior does not longer 
hold.

\begin{figure}
    \centering
   \rotatebox{0}{\resizebox{.8\textwidth}{!}{\includegraphics{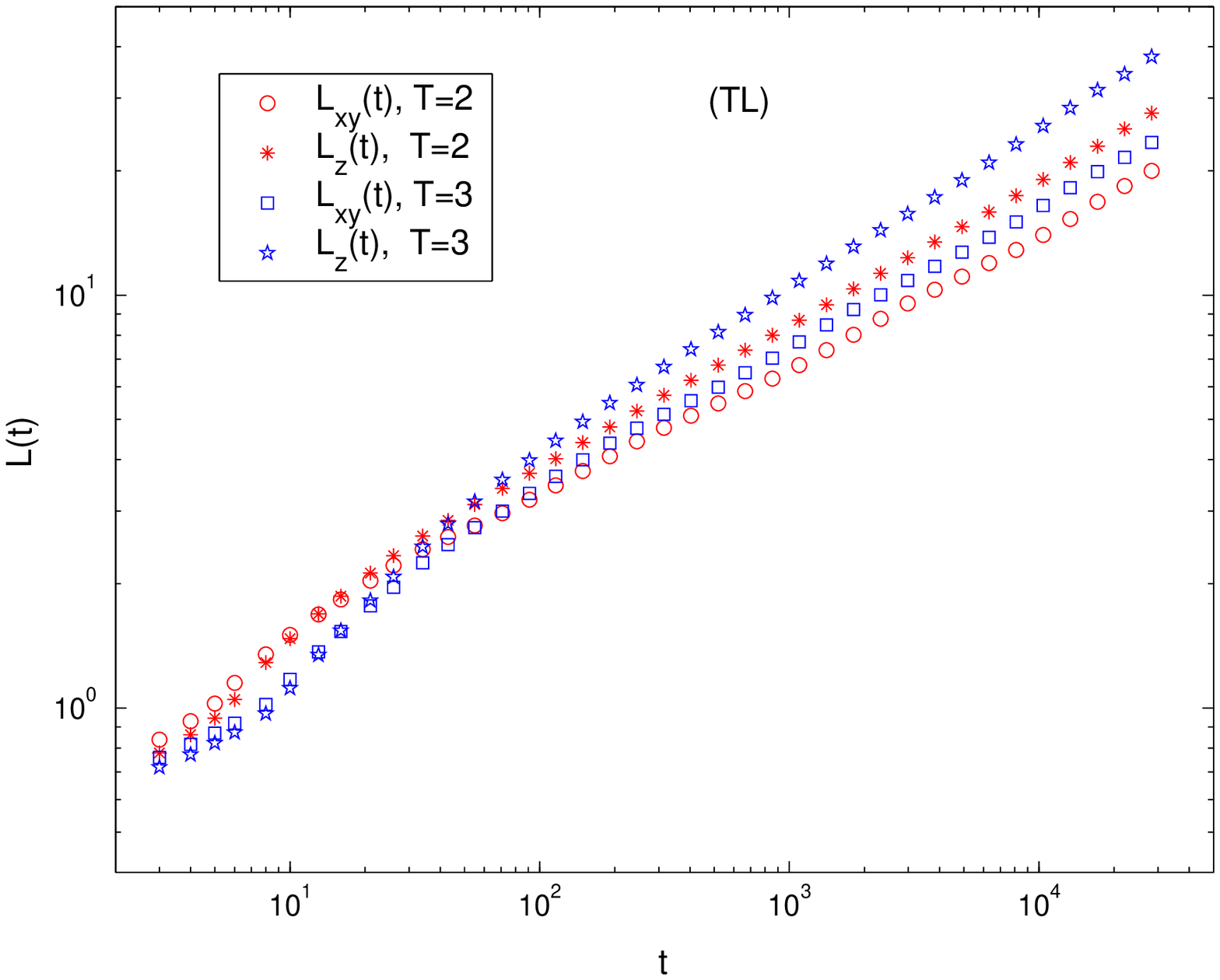} \includegraphics{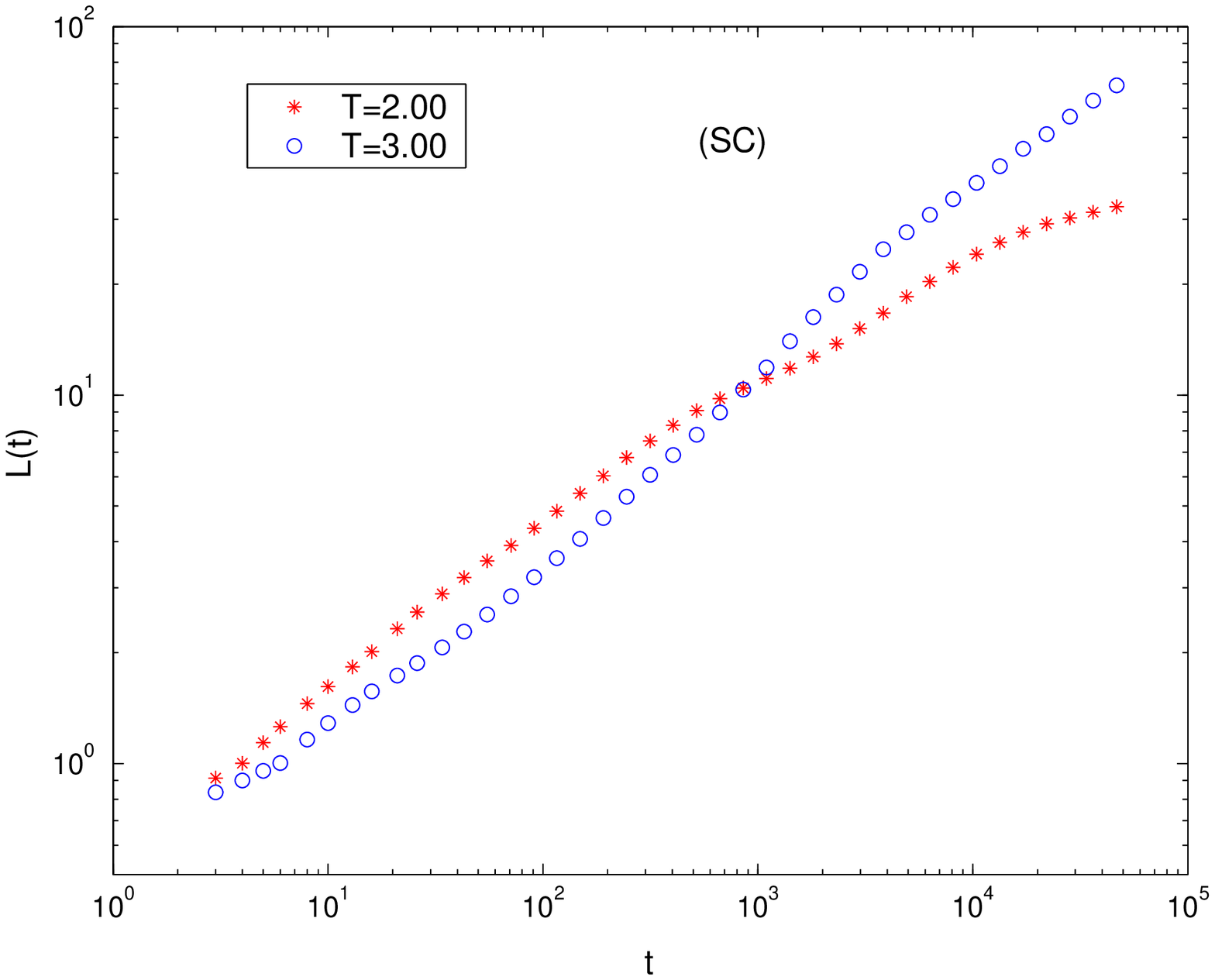}}}
    \caption{For the TL The characteristic lengths $L_{xy}(t)$ and $L_z(t)$ are 
plotted at the temperatures $T_f=2.0$ and $T_f=3.0$. For the (SC) $L(t)$ is plotted against $t$, at $T_f=2.0$ and $T_f=3.0$.}
\label{length2}
\end{figure}

\begin{figure}
    \centering
   \rotatebox{0}{\resizebox{.8\textwidth}{!}{\includegraphics{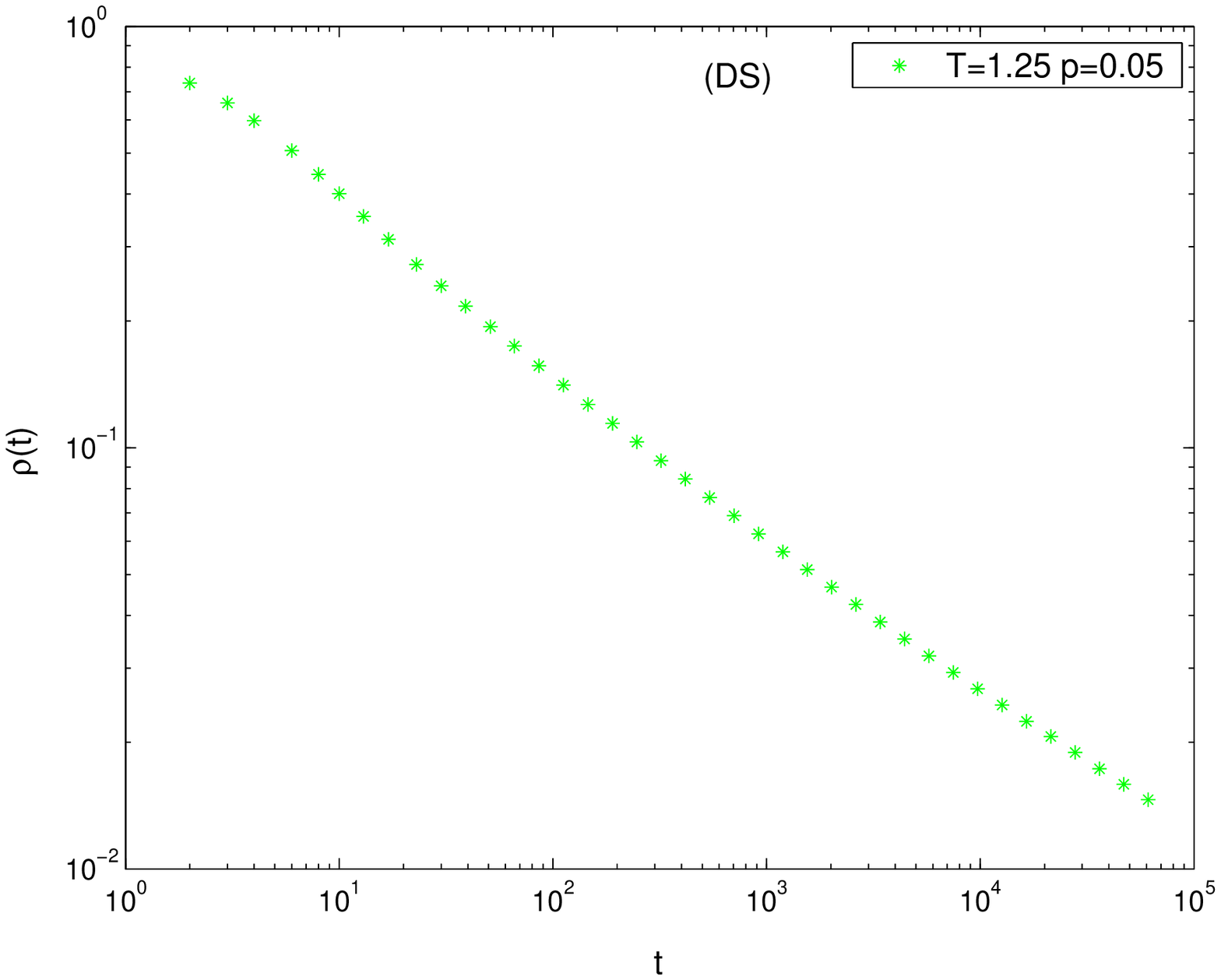}\includegraphics{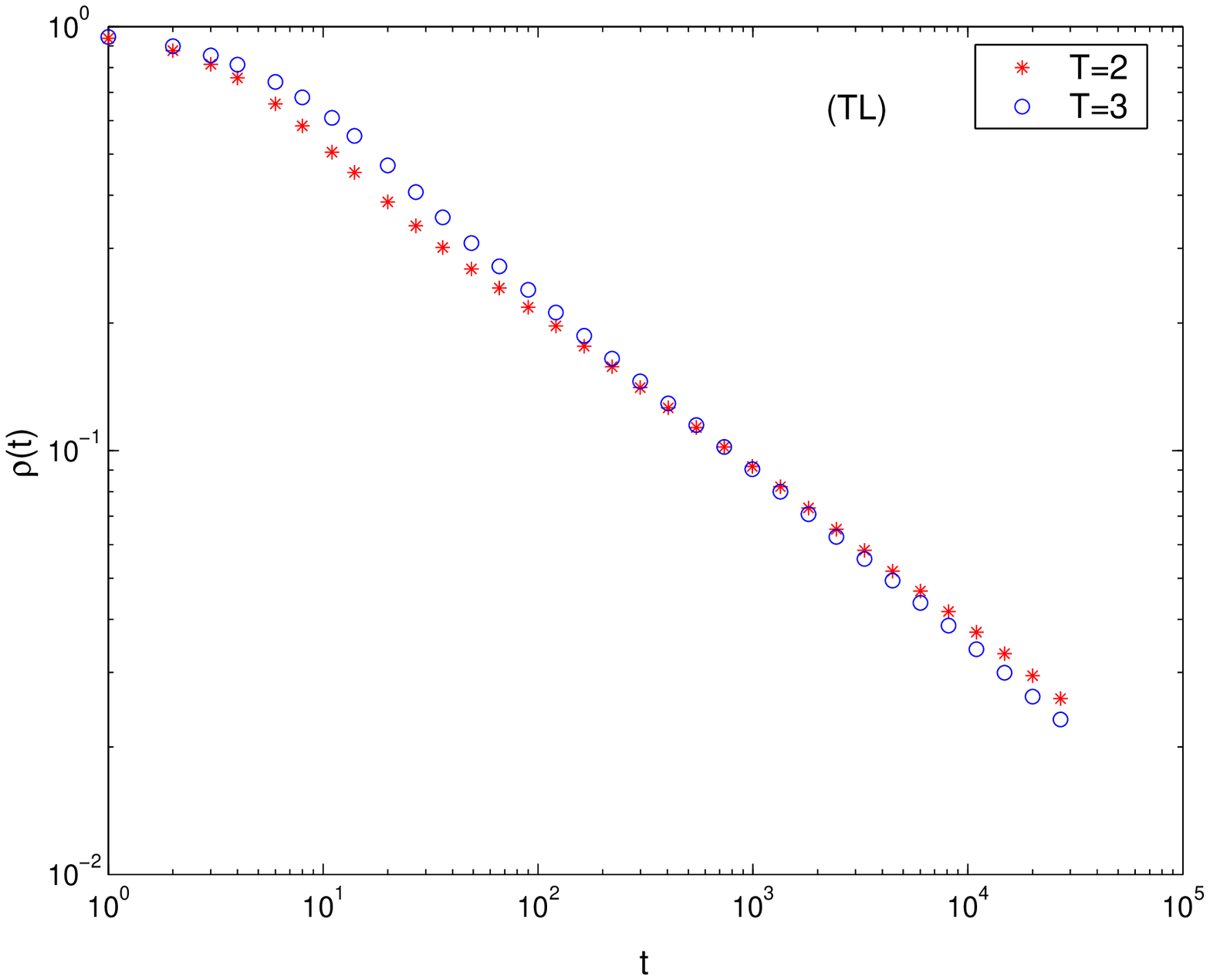}}}
 \rotatebox{0}{\resizebox{.4\textwidth}{!}{\includegraphics{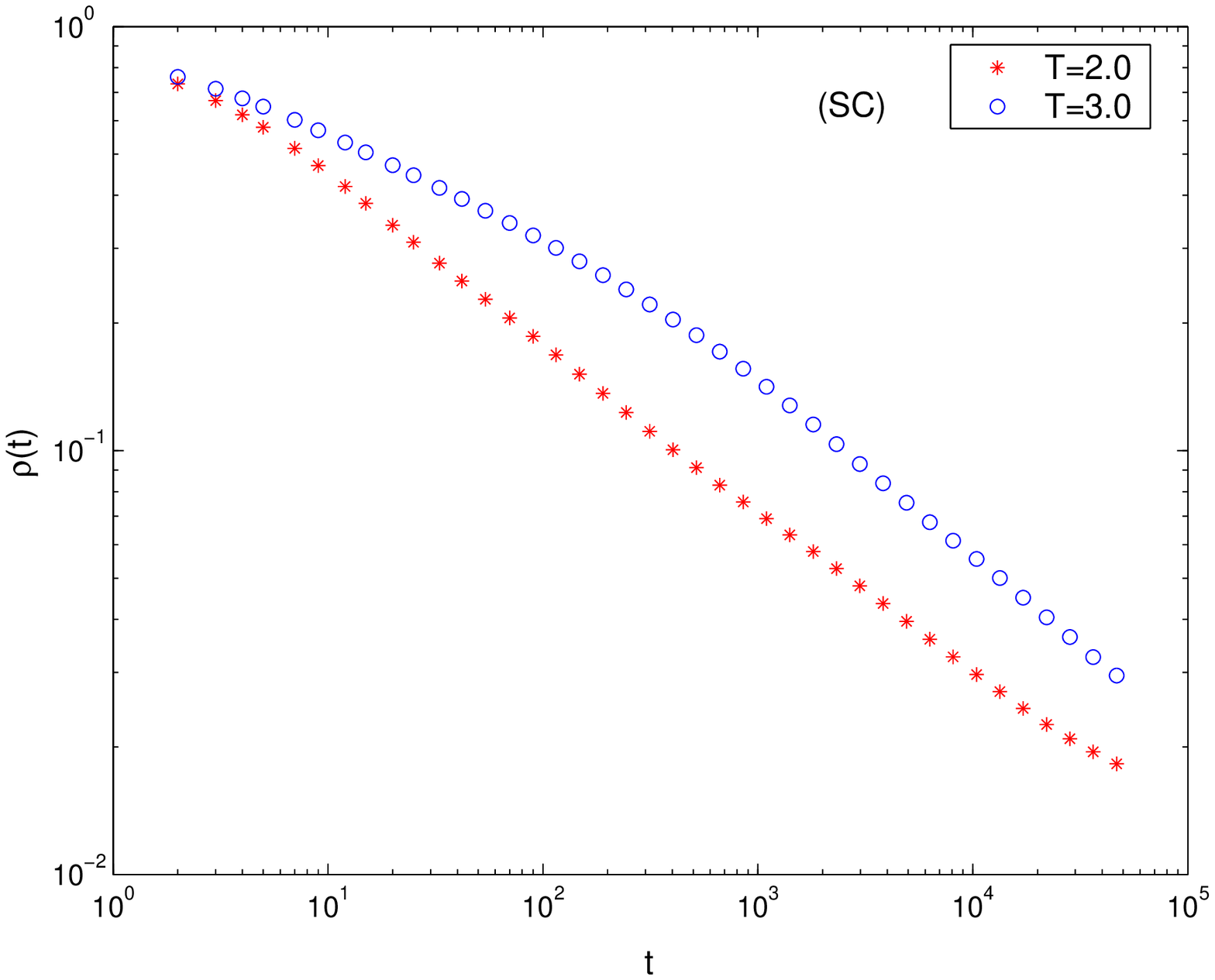}}}
    \caption{(DS) The density of interfaces $\rho (t)$ is plotted against $t$, at the temperature $T_f=1.25$ for the (DS), at $T_f=2.00$ and $T_f=3.00$ for the (TL) and for the (SC).}
\label{bordi2}
\end{figure}

The characteristic size $L(t)$, obtained as in Eq.~(\ref{halfh}), is
shown in Fig.~\ref{length2}. For the TL there are two
different lengths $L_{xy}(t)$ and $L_z(t)$ obtained from  
$G^{ag}_{xy}(r,t)$ and $G^{ag}_{z}(r,t)$ respectively. Note also in this case the presence of oscillations on top of
an average power-law growth, as discussed in Sec.~\ref{nophasetr}. For the SC the exponent $z$ is $z\simeq 0.33$ and $z\simeq 0.45$ at $T_f=2.00$ and $T_f=3.00$. The behavior of $\rho (t)$ is similar to that found for the structures 
with $T_c=0$, discussed in Sec.~\ref{nophasetr}. This quantity,
is shown in Fig.~\ref{bordi2} for the DS, TL and SC. For the SC, $\theta \simeq 0.38$ and $\theta \simeq 0.43$ 
at $T_f=2.0$ and $T_f=3.0$. Therefore also on these structures both $\theta$ and $z$ depends on the temperature.

\begin{figure}
    \centering
   \rotatebox{0}{\resizebox{.8\textwidth}{!}{\includegraphics{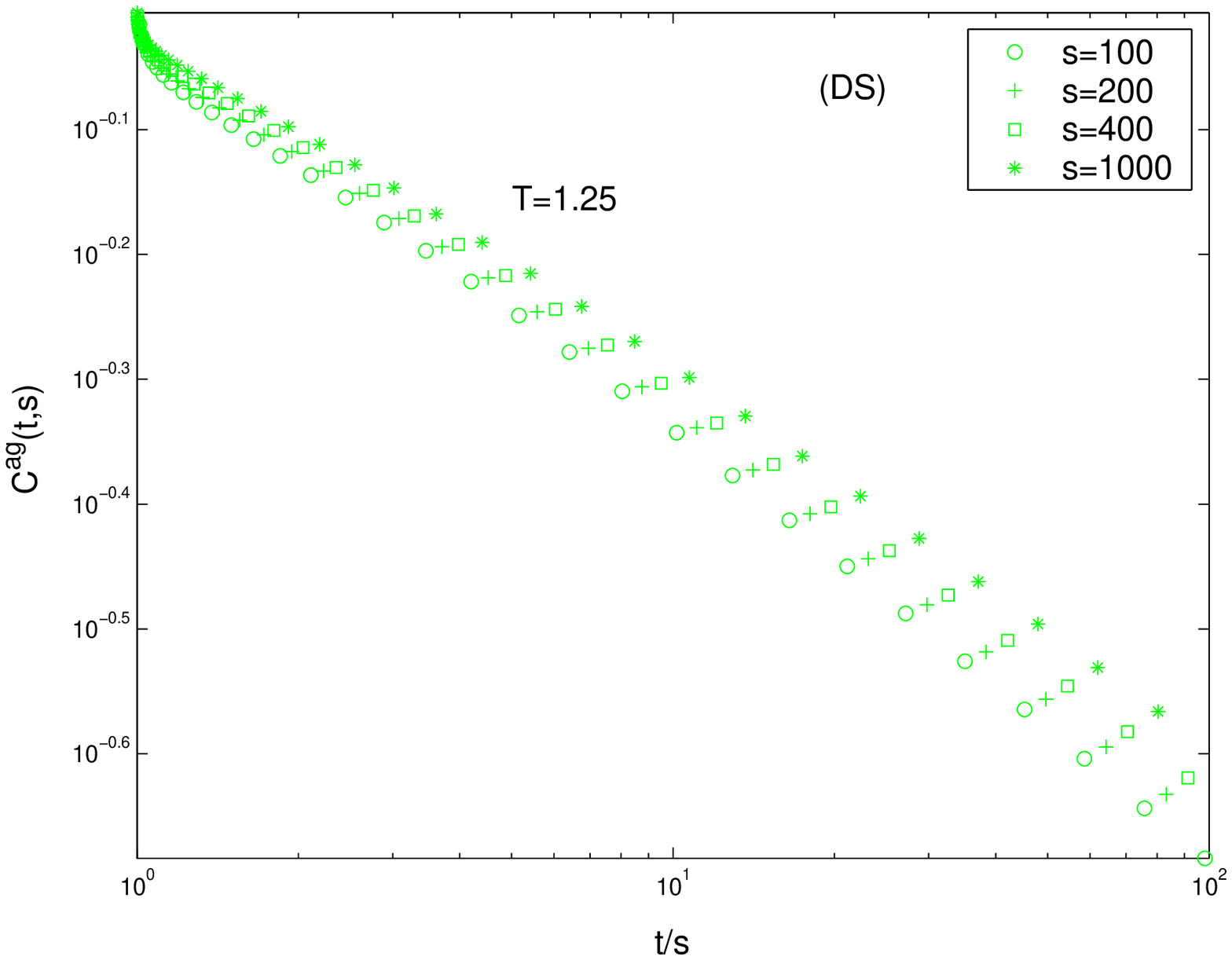}\includegraphics{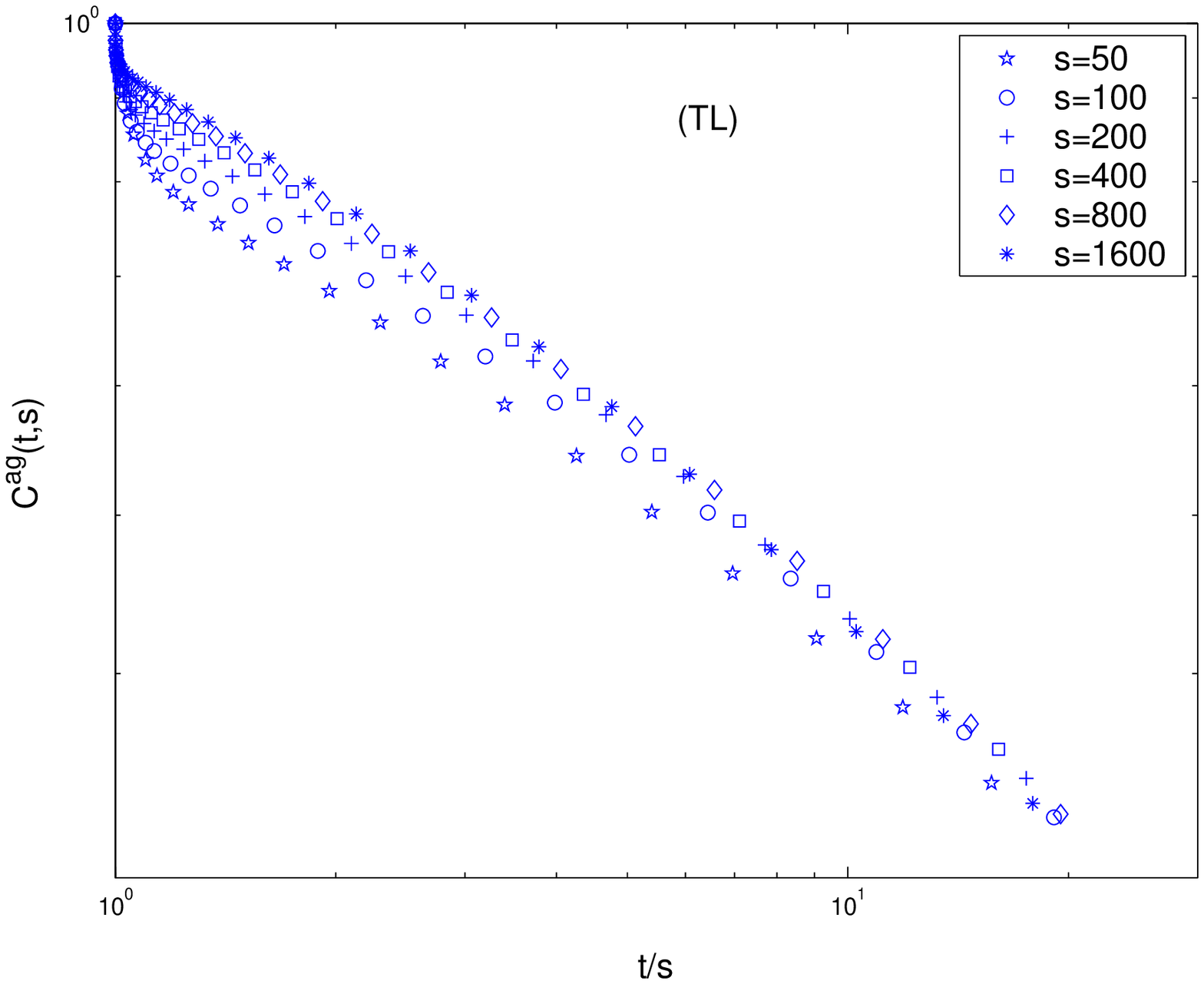}}}
\rotatebox{0}{\resizebox{.4\textwidth}{!}{\includegraphics{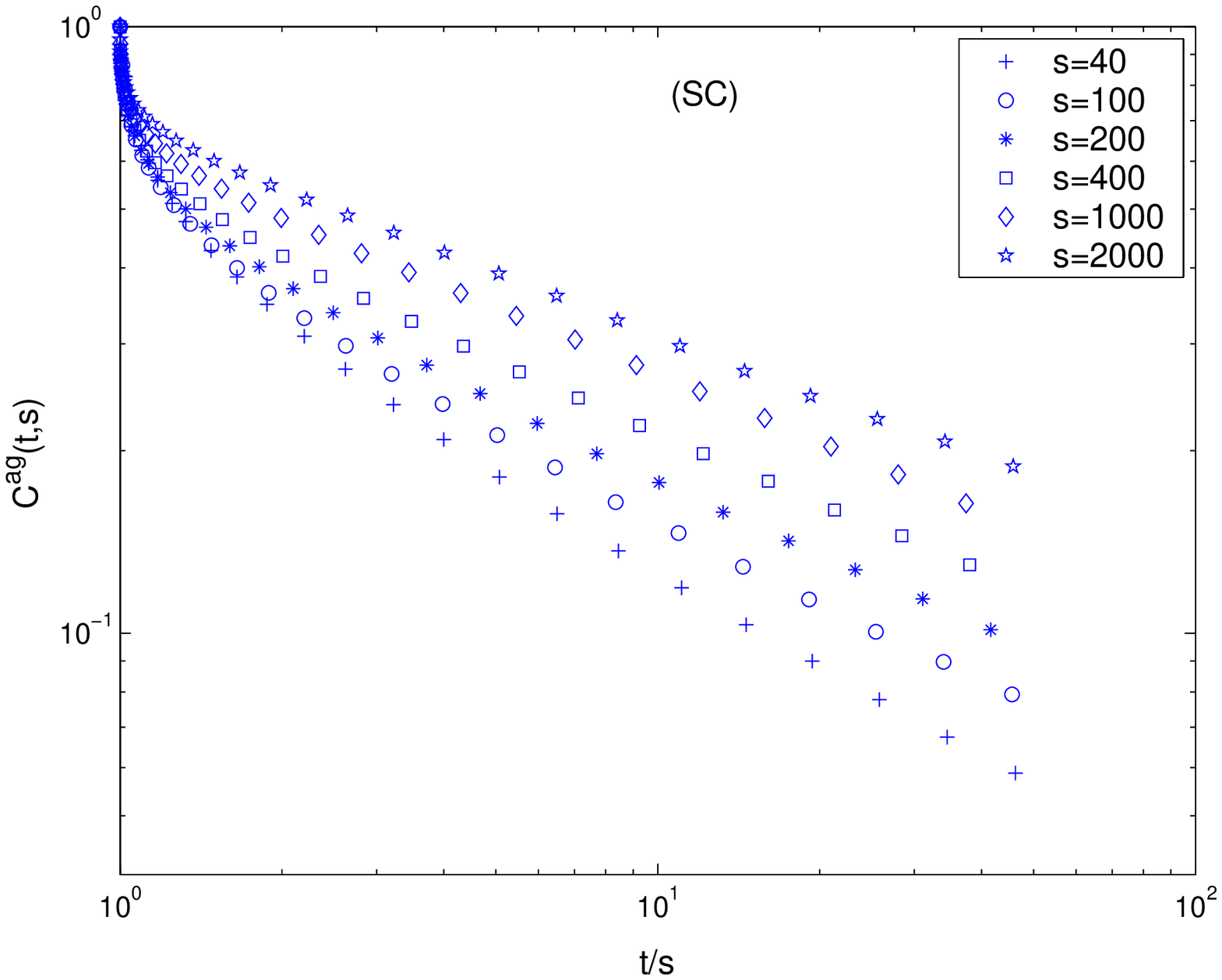}}}
    \caption{(DS) $C^{ag}(t,s)$ is plotted against $y=t/s$, at 
    $T_f=1.25$ for the (DS) and at $T_f=3.00$ for the (TL) and (SC).}
\label{scalc2}
\end{figure}

Let us consider now the behavior of two time quantities.
The autocorrelation function for the DS, TL and SC are plotted in 
Fig.~\ref{scalc2} against $y=t/s$.

For the TL one has a good data collapse for the larger values of $s$, 
according to Eq.~(\ref{scalauto}). The situation is different for the
DS and SC. In these cases, there is no data collapse in the range of time accessed in
our simulations. Probably this is due to preasymptotic effects and
larger values of $s$ then those presented in the figure should be needed
to obtain the scaling of  Eq.~(\ref{scalauto}).

Let us turn to consider the response function.
From $\chi ^{ag}(t,s)$ we extract $a_\chi $ as
the slope of a double logarithmic plot of $\chi ^{ag}(t,s)$ against $s$,
with $y$ held fixed, as described regarding 
Fig.~\ref{scalchi1tf}.

\begin{figure}
    \centering
   \rotatebox{0}{\resizebox{.8\textwidth}{!}{\includegraphics{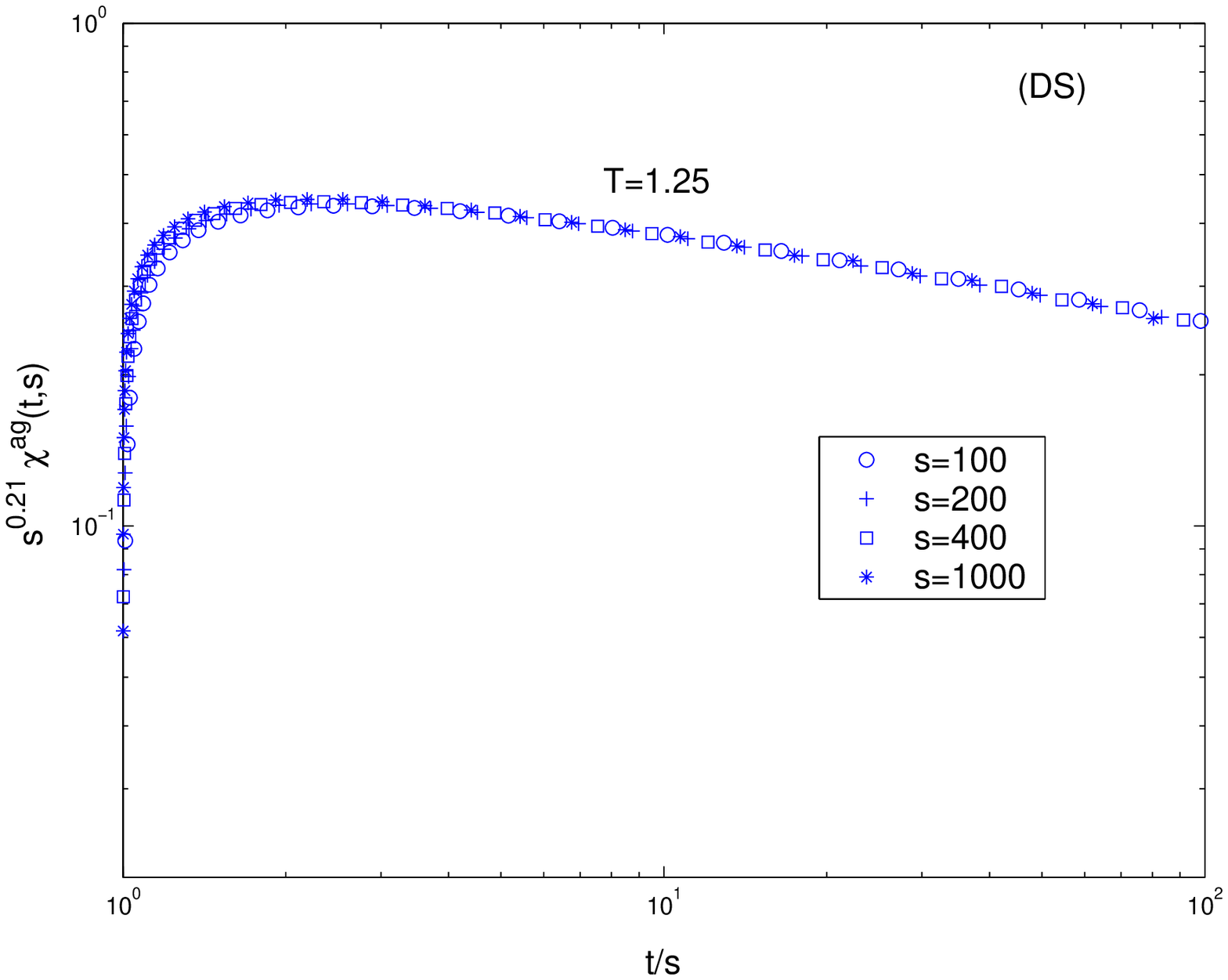}\includegraphics{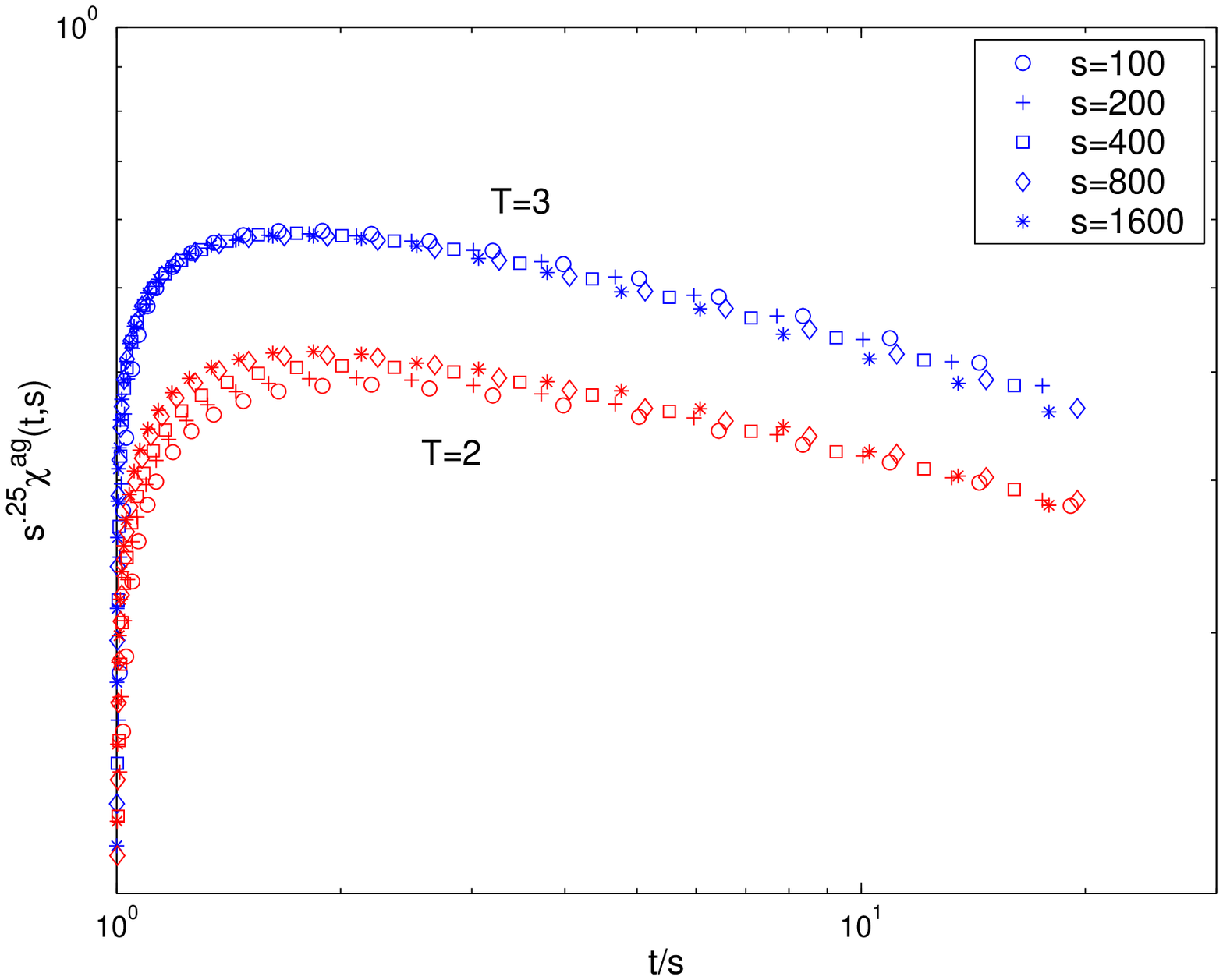}}}
\rotatebox{0}{\resizebox{.4\textwidth}{!}{\includegraphics{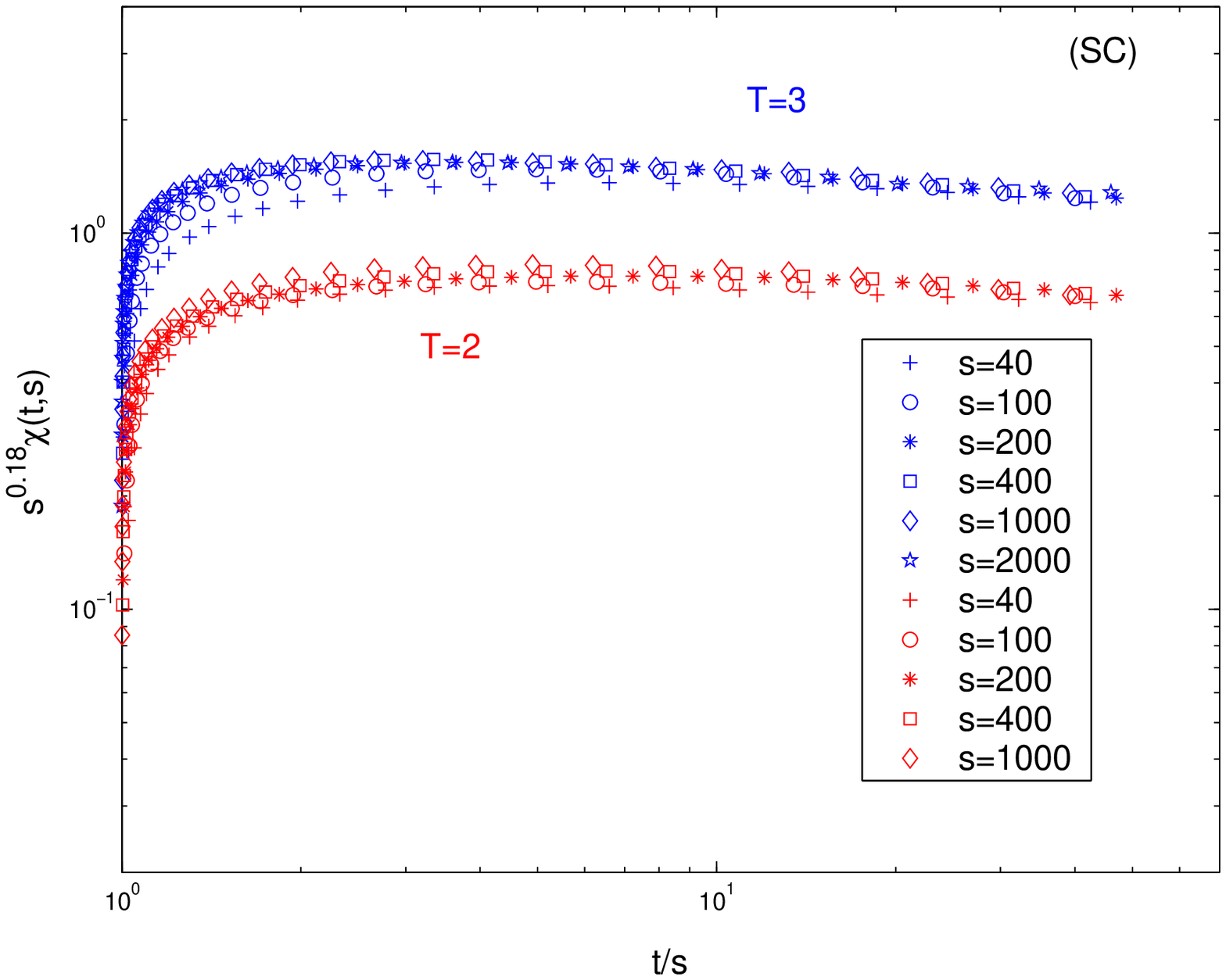}}}
    \caption{$\chi ^{ag}(t,s)$ is plotted against $y=t/s$ for different values of $s$, at    $T_f=1.25$ for the (DS), at $T_f=2.00$ and $T_f=3.00$ for the (TL) and the (SC).}
\label{scalchi2}
\end{figure}

We obtain $a_\chi =0.21$, $a_\chi =0.25$ and $a_\chi =0.18$ for the DS, the TL and the SC, respectively. 
Plotting $s^{a_\chi}\chi ^{ag}(t,s)$ against $y$ we find a good data collapse, independently from
the temperature, as shown in Fig.~\ref{scalchi2},
indicating that Eq.~(\ref{scalresp}) is obeyed.
Also the large-$y$ behavior~(\ref{largeychi}) is obeyed.

\section {The response function exponent and the fluctuation dissipation plot} \label{response}

In the previous Section, we have checked the validity of the scaling 
form~(\ref{scalresp}) in the fractal structures considered, and we have determined
the value of the exponent $a_\chi $. As already emphasized, while all the other
quantities turn out to be very sensitive to the final temperature of the quench,
this exponent assumes a well defined value in all the low temperature region.  
On regular lattices, analytical calculations in the vector ${\cal O}({\cal N})$ model
in large-${\cal N}$ limit~\cite{ninf}
find the following dependence on dimensionality
\be
    a_{\chi} = \left \{ \begin{array}{ll}
        \theta \frac{d-d_L}{d_U-d_L}  \qquad $for$ \qquad d < d_U  \\
        \theta  \qquad $with log corrections for$ \qquad d=d_U \\
	\theta   \qquad $for$ \qquad d > d_U, 
        \end{array}
        \right .
\label{aphenom}
\ee
where $d_L$ is the lower critical dimension of static critical phenomena and
$d_U=4$. Numerical simulations~\cite{altri1,altri2,Castellano04} of discrete and
 ${\cal O}({\cal N})$ vectorial systems, with conserved and non conserved order parameter, 
are consistent
with Eq.~(\ref{aphenom}), and $d_U=3$ or $d_U=4$ for scalar or vectorial models respectively. 
This shows that the non-equilibrium exponent, $a_\chi$,  
is related in a non-trivial way to the topology of the underlying lattice, through $d$. 
Interestingly, this relation implies
$a_\chi>0$ when the system is above $d_L$, i.e. when a phase transition 
at finite temperature $T_c$ is present, 
$a_\chi = 0$ at $d_L$ and $a_\chi < 0$ below $d_L$.
This result for regular lattices 
suggests that the
non-equilibrium dynamics of statistical models, and in particular
the response function exponent, 
could be related to important topological properties also in the
case of generic graphs.
This hypothesis can be checked in some detail in systems with a continuous symmetry,
because in this case it is well known that a large scale parameter, 
the "spectral dimension" $d_s$, encodes the relevant
topological features, uniquely determines the existence of phase 
transitions~\cite{fss} and controls the critical behavior~\cite{sferico}.
In other words on generic networks 
$d_s$ plays the same role 
played by the Euclidean dimension $d$ on regular lattices.
In particular, vectorial models on graphs exhibit a phase transition
at finite temperature only if $d_s>d_L=2$. 
Solving an ${\cal O}({\cal N})$ model in the large-${\cal N}$ limit \cite{prlnostro} one can show that this property holds true also for the non-equilibrium
exponent $a_\chi$. In fact, one finds 
the same expression~(\ref{aphenom}) as for regular lattices, 
with $d_s$ replacing $d$. Then, again one has
$a_\chi>0$ when the system is above $d_L$, i.e. when a phase transition 
at finite temperature $T_c$ is present, 
and $a_\chi \le 0$ when $T_c=0$ \cite{prlnostro}.
The conclusion is that, for models with a continuous symmetry there is a well defined relation
between the non equilibrium exponent $a_\chi $ and the topological
features of the network which regulate the critical properties. In this case, this
features are totally described by a single index, namely $d_s$.
The next question is if a similar picture hold for scalar systems,
where the counterpart of $d_s$ is not known. Namely, if the non equilibrium
exponent $a_\chi $ is related to the critical properties, and in turn to topology,
such that $a_\chi >0$ or $a_\chi \le 0$ holds for graphs with $T_c>0$
or $T_c=0$ respectively. An argument~\cite{prlnostro} based on the 
topology of class II networks suggests that $a_\chi =0$ is expected for these
systems.
The values
of $a_\chi$ reported in the previous Section~\ref{structures} show quite convincingly that 
the above conjecture is verified
in all the cases considered in our simulations. One has $a_\chi \simeq 0$
in all the structures of class II and $a_\chi>0$ in those of class I.                            
This result is particularly
interesting for discrete symmetry models. In fact, while for
${\cal O}({\cal N})$ models the presence of phase transition can be inferred from 
$d_s$, there is not a such a general criterion for discrete models.
Our data suggest that $a_\chi $ may be used to infer the presence of
phase transition on a generic network. 
We recall again that this result directly links $a_\chi $ to some
relevant topological features of graphs. For this reason, although $a_\chi $
is a non-equilibrium dynamical exponent, it is totally independent 
on all the non-universal details of the dynamics we have described
in Sec.~\ref{structures}. We have already observed, in fact, that,
differently from all the other dynamical exponent, its value is
robust and temperature independent.

Finally, we mention that the value $a_\chi =0$ found in the structures of
class II is associated to an anomalous fluctuation-dissipation plot~\cite{altri1}.
Re-parameterizing in $\chi (t,s)$ the time $t$ in terms of $C(t,s)$, 
according to the theorem by Franz, Mezard, Parisi and Peliti~\cite{Franz98}, 
under broad hypoteses the following relation
\be
-T_f\lim _{s\to \infty} \left . \frac {d^2 \chi (C,s)}{dC^2}\right \vert _{C=q}=P_{eq}(q)
\label{fmpp}
\ee 
exists between the non equilibrium two time functions $\chi ,C$
and the equilibrium probability distribution $P_{eq}(q)$ of the overlaps
$Q([\sigma ],[\sigma '])=(1/N)\sum _{i=1}^N \sigma _i \sigma_i'$ 
between two configurations $[\sigma ],[\sigma ']$ 
\be
P_{eq}(q)=\frac{1}{Z^2}\sum _{[\sigma ],[\sigma ']}\exp \{ -\beta (H[\sigma]+H[\sigma '])\}
\delta (Q([\sigma ],[\sigma '])-q).
\ee
Using the $\delta$-like form of $P_{eq}(q)$ of the low temperature state of
the ferromagnetic models considered in this paper, one obtains~\cite{altri1}
the well known broken line fluctuation-dissipation plot of $\chi $ versus $C$. 
However, as discussed in~\cite{prlnostro}, when $a_\chi =0$ the theorem~(\ref{fmpp})
cannot be applied and one gets a non-trivial fluctuation-dissipation plot
which is not related to $P_{eq}(q)$.
 
\section{Discussion and conclusions} \label{concl}

In this paper we have studied the phase-ordering kinetics of the Ising model with
spin flip dynamics on a class of physical graphs. The evolution is in many respects qualitatively 
similar to what observed on
regular lattices. After the quench one has the formation and growth of domains
of the two competing ordered phases, and a scaling symmetry is obeyed quite
generally (except for the SC in the timescale of our simulations). 
However, differently from coarsening on
regular lattices, the fractal nature of the networks
pins the interfaces on locally stable positions. Escape from the pinned positions
is achieved by means of activated processes, but subsequently the interfaces are
trapped again. When interfaces are pinned, the growth of the domain size $L(t)$
is inhibited. Because of this recurrent phenomenon the usual power growth-law
of $L(t)$ is modulated by an oscillation, which is more evident when the pinning
is stronger, namely at low temperatures. The necessity of activated moves on all
time/space scales makes a great difference with respect to regular lattices. 
On regular lattices the temperature is an irrelevant parameter~\cite{Bray94}, in the sense of the
renormalization group.
Namely, in all the low temperature region the dynamics is 
controlled by the zero temperature fixed point. One has, therefore, the same
dynamical exponents in all quenches to $T_f<T_c$, regardless of the temperature.
On fractal networks, instead, the situation is different, because the dynamics of activated processes
are faster the higher is $T_f$. As a consequence, exponents are always found
temperature dependent. 

A notable exception is the response function exponent $a_\chi $,
which turns out to be stable and temperature independent. 
This already suggests that $a_\chi $ may be related to a more fundamental property of
the system which remains stable under temperature variations.
Interestingly, we find $a_\chi =0$ on all the considered graphs which do not support
a ferro-paramagnetic transition at a finite critical temperature, while $a_\chi >0 $
in all the other cases. The same situation was found~\cite{ninf,altri1,altri2} 
on regular lattices,
where $a_\chi$ is related to the euclidean dimension. On these lattices,
the euclidean dimension is the topological parameter that determines
the exhistence of phase transitions and regulates the critical properties.
This calls for the hypoteses that, also on generic graphs, $a_\chi $ could be
related to the relevant topological features which govern the critical behavior,
although for systems with a discrete symmetry, such as the Ising model considered
here, a unique topological indicator, analogous to $d$ is not known.
Interestingly, this fact suggests that 
$a_\chi $ can be used to infer topological properties of graphs.

{\bf Acknowledgments} - This work has been partially supported
from MURST through PRIN-2004.

\end{document}